\documentclass[11pt,a4paper,english,amssymb,nofootinbib,superscriptaddress]{revtex4}
\usepackage{lmodern}

\usepackage[T1]{fontenc}
\usepackage[latin9]{inputenc}
\usepackage{babel}
\usepackage{amsmath}
\usepackage{graphicx}
\usepackage{amssymb}
\usepackage{esint}
\usepackage[unicode=true, pdfusetitle,
bookmarks=true,bookmarksnumbered=false,bookmarksopen=false,
breaklinks=false,pdfborder={0 0 1},backref=false,colorlinks=false]
{hyperref}
\def\b{\begin{equation}}
\def\e{\end{equation}}

\makeatletter
\@ifundefined{textcolor}{}
{%
 \definecolor{BLACK}{gray}{0}
 \definecolor{WHITE}{gray}{1}
 \definecolor{RED}{rgb}{1,0,0}
 \definecolor{GREEN}{rgb}{0,1,0}
 \definecolor{BLUE}{rgb}{0,0,1}
 \definecolor{CYAN}{cmyk}{1,0,0,0}
 \definecolor{MAGENTA}{cmyk}{0,1,0,0}
 \definecolor{YELLOW}{cmyk}{0,0,1,0}
 }

\usepackage{latexsym}\usepackage{bm}

\makeatother

\makeatother

\begin{document}

\title{Causality in 3D Massive Gravity Theories}

\author{Jos\'e D. Edelstein}
\affiliation{Department of Particle Physics and IGFAE\\
Universidade de Santiago de Compostela, E-15782 Santiago de Compostela, Spain}

\author{Gaston Giribet}
\affiliation{Physics Department, FCEyN-UBA $\&$ IFIBA-CONICET\\
Ciudad Universitaria, Pabell\'on I, 1428, Buenos Aires, Argentina}
\affiliation{Martin Fisher School of Physics, Brandeis University
\\
Waltham, Massachusetts 02453, USA.}

\author{Carolina G\'omez}
\affiliation{Physics Department, FCEyN-UBA $\&$ IFIBA-CONICET\\
Ciudad Universitaria, Pabell\'on I, 1428, Buenos Aires, Argentina}
\affiliation{Universit\`{a} degli studi di Milano Bicocca
\\
Piazza della Scienza 3, 20161, Milano, Italy.}

\author{Ercan Kilicarslan}
\affiliation{Van Swinderen Institute for Particle Physics and Gravity\\
University of Groningen, Nijenborgh 4, 9747 AG Groningen, The Netherlands}
\affiliation{Department of Physics\\
Middle East Technical University, 06800, Ankara, Turkey}

\author{Mat\'{\i}as Leoni}
\affiliation{Physics Department, FCEyN-UBA $\&$ IFIBA-CONICET\\
Ciudad Universitaria, Pabell\'on I, 1428, Buenos Aires, Argentina}

\author{Bayram Tekin}
\affiliation{Department of Physics\\
Middle East Technical University, 06800, Ankara, Turkey}
 
\date{\today}

\begin{abstract}
We study the constraints coming from local causality requirement in various $2+1$ dimensional dynamical theories of gravity. In topologically massive gravity, with a single parity non-invariant massive degree of freedom, and in new massive gravity, with two massive spin-$2$ degrees of freedom, causality and unitarity are compatible with each other and both require the Newton's constant to be negative. In their extensions, such as the  Born-Infeld gravity and the minimal massive gravity the situation is similar and quite different from their higher dimensional counterparts, such as quadratic ({\it e.g.}, Einstein-Gauss-Bonnet) or cubic theories, where causality and unitarity are in conflict. We study the problem both in asymptotically flat and asymptotically anti-de Sitter spaces. 

\end{abstract}

\maketitle

\section{Introduction}
Shapiro's time-delay argument \cite{Shapiro:1964uw}, known as the fourth test of general relativity (GR), basically says that light making a round-trip in space takes the least time in the absence of gravity, that is in Minkowski space. This is true in GR as can be demonstrated in several ways \cite{Gao:2000ga}. But it is not automatically the case in gravity theories having quadratic or cubic curvature terms, where causality violation ---ultimately attributed to a Shapiro time advance--- was noticed \cite{Camanho:2014apa}. Interestingly enough, causality restoration demands the inclusion of an infinite tower of massive higher-spin particles with fine-tuned interactions that imply Reggeization \cite{Camanho:2014apa,D'Appollonio:2015gpa,Camanho:2015bbt}, an apparently distinctive fingerprint of perturbative string theory.

Here we explore the status of this problem in $2+1$ gravity. Naively one might think that the Shapiro time-delay does not play any role in $2+1$ dimensions, given the standard lore stating that there are no local gravitational degrees of freedom. However, there are several metric-based massive dynamical, locally nontrivial  gravity theories in $2+1$ dimensions that have received quite a lot of attention in the earlier and recent literature, and we want to scrutinize on them under the light of causality. Since, of course, the $2+1$ dimensional scenario is not open to real experiments, the main question is to understand whether the {\it causality} and {\it unitarity} conditions are in contradiction or not. By unitarity, we mean the absence of ghosts and tachyons in the linearized excitations about the vacuum of the theory, and by causality we mean the positivity of the time-delay ---as opposed to a time-advance---, along the line of Shapiro's argument. 

We will only consider the local causality problem and not get involved with more complicated global causality issues, as even the four dimensional GR is also not known to be immune to them, at least in the presence of matter, albeit unphysical as is well-known in the canonical example of the G\"{o}del spacetime. For this purpose, it is sufficient to consider spaces that asymptote to maximally symmetric backgrounds. We will first consider asymptotically Minkowski solutions; we will do so in the first part of the paper. In the second part, we will study the case of asymptotically Anti-de Sitter (AdS) spaces. 

In three dimensions, Einstein's gravity has no local propagating degrees of freedom, and hence there is no issue of local causality; global causality issues are dealt with in \cite{thooft}. In other words, in pure Einstein's gravity in 3 spacetime dimensions all the solutions are locally equivalent to Minkowski, de Sitter, or Anti-de Sitter space, depending on whether the cosmological constant is zero, positive, or negative, respectively. The Riemann curvature is constant, except for the conical defects associated to the mass distribution; the theory has no propagating gravitons, and there is no room for Shapiro time-delay whatsoever. In contrast to 3-dimensional general relativity, topologically massive gravity (TMG) \cite{DJT} as well as new massive gravity (NMG) \cite{BHT} do have local massive propagating modes, and in these theories the discussion about unitarity reduces to that of the correct choices for the signs of the kinetic and mass terms of the linearized excitations. A priori, as in the case of Einstein-Gauss-Bonnet or cubic theories in higher dimensions, there is the possibility that causality and unitarity are in conflict with each other leading to a physically troublesome theory or prompting the conclusion that the theory is at best an effective one. But in TMG, NMG and their modifications, we will show that once the sign of the Einstein-Hilbert term in the action is chosen to be the one required for unitarity --namely, the opposite to that of the $3+1$ dimensional case-- there is no conflict with local causality. Reversing the sign of the Einstein-Hilbert term basically is equivalent to taking the $2+1$ dimensional Newton's constant, $G$, to be negative. It remains somewhat a puzzle as to why $2+1$ dimensions is different in this sense from all the higher dimensional cases.

There are at least two ways to calculate the Shapiro's time-delay: The usual method is to consider the black hole solutions and look at the time-delay of light in a round-trip in the presence of the black hole
compared to the absence of it. Another way is to calculate the time-delay of a massless particle moving in the presence of a shock wave \cite{shock, Dray} created by a high-energy massless particle. The second method is better suited to the $2+1$ dimensions in asymptotically flat space, since in that case black hole solutions are not available in TMG and NMG and other massive gravity theories --except for the case of purely quadratic theories--.

After studying the problem in flat space, we will address the case of negative cosmological constant. In the last part of the paper, the question we will address is whether the same phenomenon occurs in asymptotically AdS$_3$; that is, whether a region of the parameter space exists for which the massive gravity in AdS$_3$ can be free of ghosts and tachyons and, at the same time, compatible with local causality. This question is not redundant since the new scale given by the curvature of AdS$_3$ introduces differences with respect to flat space. In fact, as we will see, there are two quantitative modifications suffered by the Shapiro time-delay relative to Minkowski space: On the one hand, the Yukawa type dependence on the impact parameter found for this kind of process in Minkowski space suffers a correction in the effective mass $m_g$, which in AdS$_3$ happens to depend not only on the graviton mass parameter $m$ but also on the cosmological constant $\Lambda $. On the other hand, the Shapiro time-delay gets multiplied by an overall factor $N$ that is a function of the Reynolds number $m_g/\sqrt{|\Lambda |}$ that tends to 1 in the limit $\Lambda \to 0$. The latter modification is important since, in principle, it means that the sign of the time-delay in AdS$_3$ could depend on the interplay between the two scales $m_g $ and $\sqrt{|\Lambda |}$.  

There is another crucial difference between flat and AdS$_3$ spaces, which is the aforementioned existence of black holes \cite{BTZ, BTZ2}. As we have previously said, in three dimensions black holes only exist\footnote{More precisely, in three dimensional massive gravity there also exist black holes in asymptotically flat and asymptotically de Sitter spaces \cite{NMG2, OTT} and in other spaces \cite{Lifshitz}; however, those solutions only exist at very special points of the parameter space where the theories exhibit special properties \cite{GGI, birkhoff}.} in AdS space, and it turns out that the sign of $G$ that renders the theory free of ghosts is the opposite to the one that makes the mass spectrum of the black holes positive-definite. This introduces an additional puzzle that, in order to be solved, demands invoking a kind of superselection argument\footnote{Other solutions have been proposed, such as looking for special values of the parameters and suitable boundary conditions that make the theory to lose its local degrees of freedom and enable one to choose the positive sign in the Einstein-Hilbert action \cite{CG}.} \cite{Superselection}, cf. \cite{Sachs}. For our purpose, this means that in order to compute the Shapiro time-delay in 3 dimensions it would not be natural to resort to the black hole geometry calculation analogous to the one usually employed in 4 or more dimensions, since what we actually want to analyze here is whether the delay is positive for the same sign of the Einstein-Hilbert action that renders the theory free of ghost-free. Therefore, the way to address the problem in AdS$_3$ will be, again, by considering a {\it gedankenexperiment} analogous to the one considered in flat space \cite{Camanho:2014apa}. We will compute the time-delay suffered by a particle moving in the presence of a shock-wave generated by a high-energy massless particle. Adapting such an experiment to the case of AdS$_3$ space requires to consider the gravitational wave solutions found in \cite{Ayon, Bending} coupled to the particle source. We will find that a particle interacting with such a shock-wave in AdS$_3$ experiences a time-delay that is positive-definite. We will see that this happens both for TMG and for NMG. 

Besides, since the main motivation for considering AdS$_3$ space comes from AdS/CFT \cite{Malda}, raising the question of the compatibility between unitarity and causality in the bulk unavoidably leads to also ask about the unitarity in the boundary conformal field theory (CFT). As it is well known, both TMG and NMG suffer from what is known as the bulk/boundary unitarity clash; that is, the discrepancy between the sign of the Einstein-Hilbert action that makes the theory free of ghosts and the one that yields a positive central charge in the dual CFT. This problem has not been solved yet, and remains one important issue in AdS$_3$ massive gravity \cite{Nuevo}. Nevertheless, an interesting attempt to solve it has led to the discovery of an interesting new type of 3-dimensional massive model, known as minimal massive gravity\footnote{It has been recently observed in \cite{Nuevo} that MMG in the metric formulation exhibits a logarithmic mode that can spoil unitarity in the bulk. The question remains as to whether the definition of the theory can be supplemented with a specification of asymptotic boundary conditions that accomplish to decouple the logarithmic mode and render the theory unitary.} (MMG) \cite{berk}. This model consists in augmenting the TMG field equations with additional second order terms that, even when do not give relevant contributions to the effects we want to investigate around flat space, do contribute in AdS$_3$ space. Therefore, we will discuss this more general in the last part of the paper, showing here that also in MMG the Shapiro time-delay turns out to be positive-definite when the MMG corrections are taken into account.

The lay-out of the paper is as follows: In the next section, we warm up by showing in the context of 3 dimensional GR the kind of computations we will use throughout the work. In section III and IV we study TMG and NMG cases  wherein we discuss the time-delay for null geodesics, for massless scalar fields, massive non-minimally coupled photons and the gravitons of the relevant theory.  In each section, we also derive some of the results from the point of view of eikonal scattering amplitudes. We also briefly discuss the case of Born-Infeld gravity in 3 dimensions. In section V we will extend the analysis to asymptotically AdS$_3$ spacetimes. After the Conclusions section, we assist the reader with various appendixes including conventions and the derivation of the relevant tensor perturbations about the shock-wave background. 

\section{General relativity warm up in 2+1 dimensions}
\label{sec2}

It is a well known fact that 2+1 GR is very different from higher dimensional GR. If we turn off the cosmological constant, the vacuum field equations imply the vanishing of every component of the Riemann tensor ---outside sources, space-time is trivially flat. As a consequence, a careful analysis of linearized perturbations around flat space shows that they can be only pure gauge: there are no local propagating degrees of freedom. Nevertheless, we will take GR as a warm up exercise to settle the structures and type of computations we shall make to analyze the causality problem in other theories of 2+1 gravity. 

Consider in GR a shock-wave ansatz\footnote{See Appendix for more details.}
\begin{equation}
ds^2=-du dv+H(u,y) du^2+dy^2 ~,\label{ansatz}
\end{equation}
sourced by an energy-momentum tensor $T_{uu}=\lvert p\rvert \delta(y)\delta(u)$ which corresponds to a massless point particle moving in the $+x$ direction with 3 momentum as $p^\mu=\lvert p\rvert(\delta^\mu_0+\delta^\mu_x)$, where $u=t-x$ and $v=t+x$ are light-cone coordinates and $y$ is the transverse coordinate. Einstein equation for $H$ becomes (we set $|8\pi G| = 1$)
\begin{equation}
\begin{aligned}
\partial_y^2H(u,y)=-2\sigma\lvert p\rvert \delta(y)\delta(u) ~, \qquad \sigma := {\rm sign}\,G ~.
\end{aligned}
\end{equation}
The most general solution to this equation is
\begin{equation}
H(u,y)=-2\sigma\lvert p\rvert\delta(u)\theta(y)y\,+\,c_1(u)y+c_2(u) ~,
\end{equation}
where $c_1(u)$ and $c_2(u)$ are arbitrary and related to the coordinate transformations which leave the ansatz invariant. Notice that if we choose $c_1(u)=0$ and $c_2(u)=0$ we would get a vanishing profile for $y<0$ but a nontrivial one for $y>0$. On the opposite hand if we choose $c_1(u)=2\lvert p\rvert\delta(u)$ and $c_2(u)=0$ we would get
\begin{equation}
H(u,y)=2\sigma\lvert p\rvert\delta(u)\theta(-y)y ~,\label{profileGR}
\end{equation}
meaning a vanishing profile for $y>0$. Actually, and consistently with what was explained before, outside the source, the space is flat everywhere, but, due to the source we cannot have a single chart to write down the metric in such a way as to have Cartesian coordinates in both sides of the profile.  Notice that a slightly different shock-wave profile was found in \cite{desershock} which agrees with the general form (3) but with constants $c_1$ and $c_2$ chosen such that the profile is symmetric under $y \rightarrow -y$. The reason we do not choose such a symmetric solution is that since our main purpose is to compute the time-delay as measured by an asymptotic observer, we pick $c_1$ and $c_2$ such that for $y>0$ we have asymptotically flat and Cartesian coordinates. In the GR case this means that the profile is trivial for $y>0$ but in more general cases a nontrivial profile will be found.

Our interest will be to consider in different theories a particle traversing the shock-wave profile for a fixed value of the impact parameter $y=b>0$. In every case, we will fix the coordinates of the profile in such a way that for $y\rightarrow\infty$ the coordinates are asymptotically flat and Cartesian. 

In particular, for the present case of GR, this choice coincides with the profile (\ref{profileGR}) and trivially, a massless spinless particle traversing this profile with $y=b>0$ will not experience any delay. While this conclusion is trivial in GR, in more general theories, we will study the discontinuity of the geodesic of the particle at $u=0$ to obtain the time-delay.  Notice that not having chosen the symmetric profile of \cite{desershock} does not mean that the existence of the delay depends on which side of the shockwave the test particle goes through. As mentioned before, space is flat at both sides of the profile and our choice of $c_1$ and $c_2$ is just a coordinate choice fixed by our utilitarian reasoning of working with asymptotically flat and Cartesian coordinates on one side ($y>0$).

Besides the geodesic analysis we will also use scattering amplitudes to confirm some of the results we obtain for the delay \cite{thooft2}. Instead of thinking of the experiment as a massless particle traversing the shock-wave geometry produced by another particle, we consider the tree-level $2\rightarrow 2$ scattering amplitude of massless non-self-interacting gravitating particles  in the deflectionless limit  $\tfrac{t}{s}\rightarrow 0$, which in the case of GR is given by
\begin{equation}
\mathcal{A}_{tree}(s,t)=-\sigma \frac{s^2}{t} ~,
\end{equation}
where $\sqrt{s}$ is the center-of-mass energy and $\sqrt{-t}$ is the absolute value of the momentum transfer. The full amplitude in the eikonal approximation will (in many practical cases) exponentiate in the impact parameter space \cite{kabat,giddings}. The phase associated to this exponentiation is proportional to the Shapiro time-delay and it can be computed by Fourier transforming the tree level result
\begin{equation}
\delta(b,s)=\frac{1}{2s}\int\frac{dq}{2\pi}\,e^{i q b}\mathcal{A}_{tree}(s,-q^2)=
\frac{\sigma}{4\pi}\,s \int dq\, \frac{e^{iqb}}{q^2} ~.\label{phaseGR}
\end{equation}
This last result diverges for the region of integration close to forward-scattering (zero momentum transfer) while we would have expected it to be zero in this GR computation. This is an artifact of the eikonal approximation where the zero angular momentum mode does not behave well in the continuous limit in the impact parameter space. We will take a pragmatic approach on this issue: we choose to regulate in some reasonable way the Fourier transform of the amplitude in such a way that the particular integral (\ref{phaseGR}), corresponding to GR, is zero as expected. In this way, by using the same prescription in other theories, the result we obtain is gauged by the GR result. 

A useful prescription is therefore to take the $q$ integral domain to be the real line of the complex $q$-plane, with an indentation of the line so as to avoid the point $q=0$ leaving the pole out of the contour of integration; say, $q = -i\epsilon$.
\begin{figure}[h]
\includegraphics[width=0.31\textwidth]{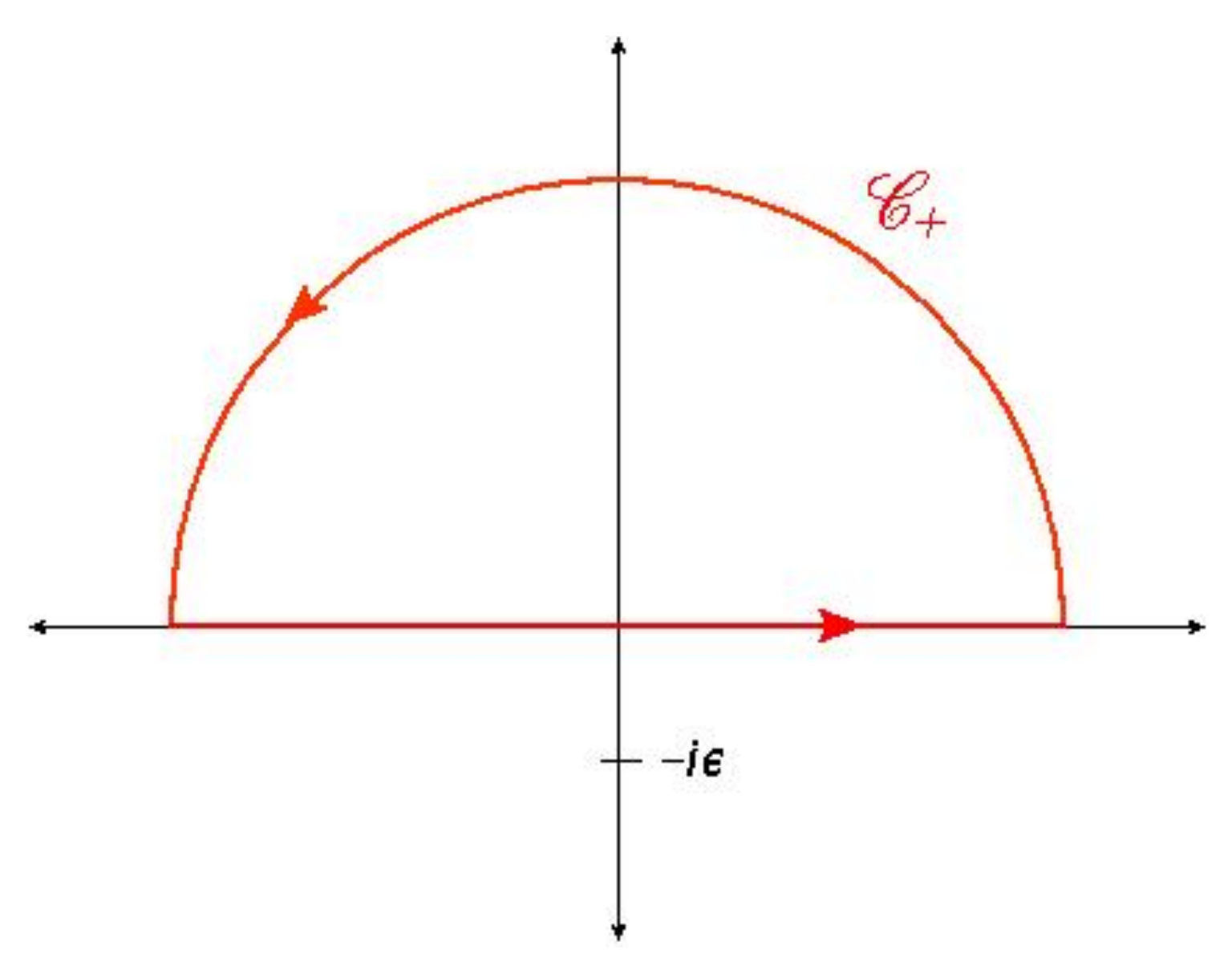}
\hspace{1cm}
\includegraphics[width=0.31\textwidth]{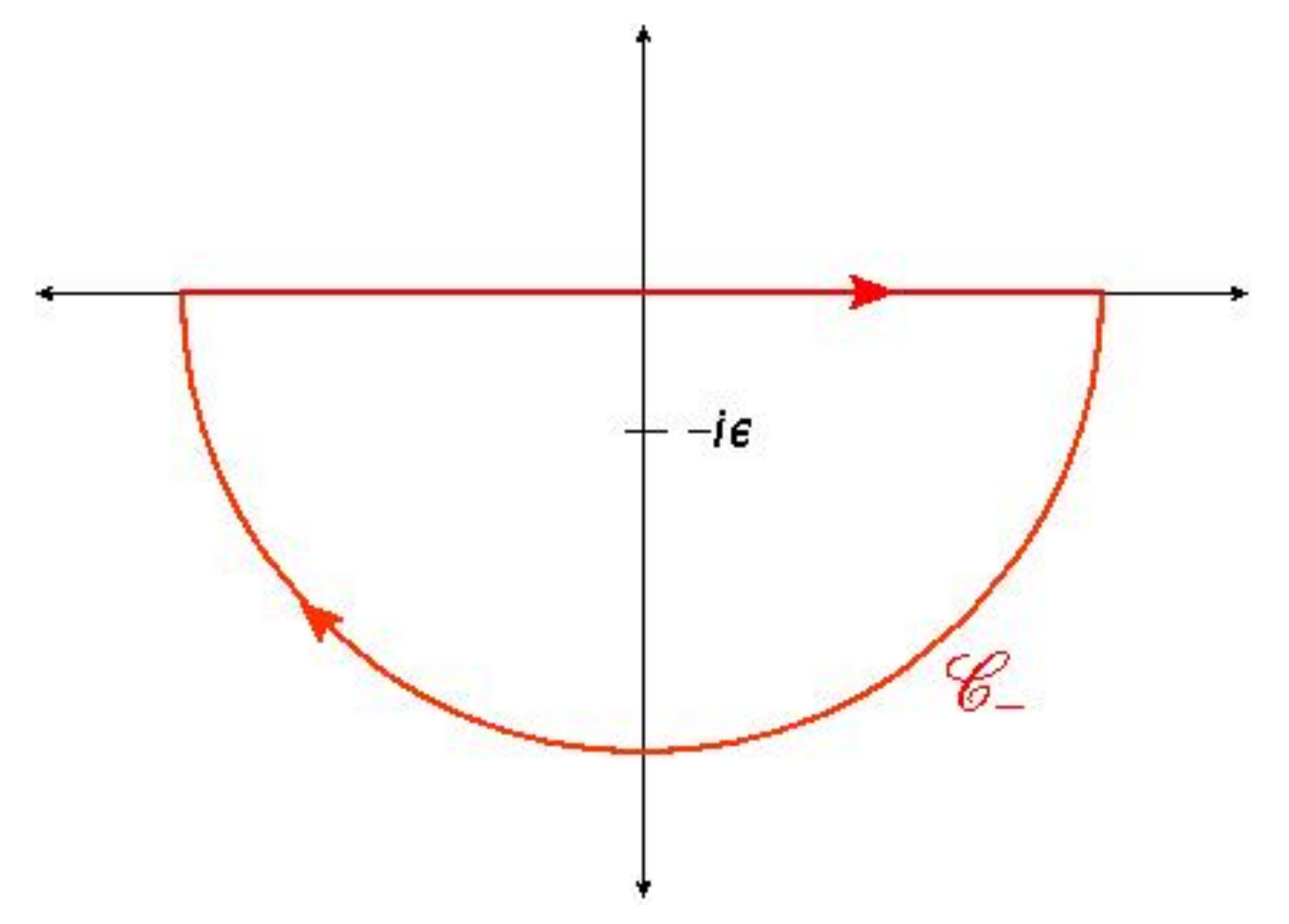}
\caption{Integration contours in the complex $q$-plane depending upon the sign of the impact parameter $b$, this prescription corresponding to the gauge choice $c_1(u)=2\lvert p\rvert\delta(u)$ and $c_2(u)=0$.}
\label{contorno}
\end{figure}
This prescription reproduces the result obtained from the geodesic analysis, provided we choose the gauge $c_1(u)=2\lvert p\rvert\delta(u)$ and $c_2(u)=0$. Displacing the pole to $q = i\epsilon$, instead, would have corresponded to the choice $c_1(u)=c_2(u)=0$, whereas the dimensional continuation of the higher dimensional result corresponds to $c_1(u)=\sigma\lvert p\rvert\delta(u)$ and $c_2(u)=0$, and leads to a symmetric profile under $y \to -y$.

We shall see later that the above discussed prescription allows us to compute the correct Shapiro time-delay for scattering scalar massless particles in other theories of 2+1 gravity besides GR. A different prescription would correspond to a choice of boundary conditions that, albeit perfectly valid, are not suitable for the sake of comparison with the Shapiro time-delay computation.

\section{Causality in TMG}
The Lagrangian density of TMG  \cite{DJT}
\begin{equation}
{\cal L}= \sqrt{-g} \, \Bigg ( \sigma R +\frac{1}{2 \mu} \, \epsilon^{\mu\nu\alpha} \Gamma^\beta{_{\mu \sigma}} \Big (\partial_\nu \Gamma^\sigma{_{\alpha \beta}}+\frac{2}{3} \Gamma^\sigma{_{\nu \lambda}}  \Gamma^\lambda{_{\alpha \beta}} \Big ) \Bigg ) ~,
\label{wtics1}
\end{equation}
when coupled to matter, yields the field equations  
\begin{equation}
\sigma G_{\mu\nu}+\frac{1}{\mu} C_{\mu\nu}=T_{\mu\nu} ,
\label{tmg}
\end{equation}
where $G_{\mu\nu}=R_{\mu\nu}-(1/2)Rg_{\mu\nu}$ is the Einstein tensor and $C_{\mu\nu}= (\epsilon_{\mu }^{\ \rho \sigma }/\sqrt{-g})\nabla _{\rho }(R_{\sigma\nu}-(1/4)Rg_{\sigma\nu})$ is the Cotton tensor. The latter is traceless, expressing the conformal invariance of this particular higher-derivative deformation of Einstein equations. We have set the Newton's constant to unity but allowed a possible sign reversal parameter $\sigma$, as discussed in the previous section, $\sigma^2 =1$.  The theory is parity non-invariant and the single helicity-2 excitation about the flat background has a mass
\begin{equation}
m_{g}=-\sigma\lvert \mu\rvert ~.
\label{masstmg}
\end{equation}
Therefore we set $\sigma =-1$  for this mode to be  non-tachyonic. This choice also ensures the kinetic energy to be positive (or non-ghostlike). Note that $\mu \to -\mu$ is a parity change, keeping the mass intact 
but reversing the helicity of the graviton. Using formulas (\ref{ricci_shock}) and (\ref{cotton_shock}) in appendix A, the TMG equations (\ref{tmg}), for the null-source, reduce to a single third derivative (carrying the burden of parity violation) equation
\begin{equation}
\begin{aligned}
-\frac{\sigma}{2}\partial_y^2 H(u,y)+ \frac{1}{2\mu}\partial_y^3 H(u,y)=\lvert p\rvert \delta(y)\delta(u) ~.
\label{eqn}
\end{aligned}
\end{equation}
Without loss of generality, let us consider  the $\mu >0$ case. (The discussion is similar, for the other sign choice without a change in the physical consequences.) Then the most general solution is easily found as
\begin{equation}
H(u,y)=-\frac{2\sigma\lvert p \rvert}{m_g}  \delta(u)  \theta (y)\bigg(e^{-m_g y}+m_g y-1\bigg) +c_1\frac{e^{-m_g y}}{m_g^2}+c_2 y+c_3 ~,
\end{equation}
with three $c_i$ functions depending on the null coordinate $u$ in an arbitrary way.  But these functions can be fixed by requiring the spacetime to be asymptotically flat.  We can use coordinate transformations 
to bring the metric to the Cartesian form in the asymptotic limit. But, one cannot bring the metric to the Cartesian form for $y>0$ and $y<0$ with a single transformation \cite{desershock}. 
This is a minor technical issue which we shall avoid and demand that for $y \rightarrow +\infty$ the space is given in terms of Cartesian coordinates, but for $y \rightarrow -\infty$, it is flat  written in non-Cartesian coordinates. (Namely, a single chart does not cover the whole spacetime as we discussed for the GR case.)  Then for $\mu>0$, 
the gauge-fixed shock wave profile of TMG reads
\begin{equation}
\begin{aligned}
H(u,y) =-\frac{2\sigma}{m_g} \lvert p \rvert \delta(u)\theta(y) e^{-m_g y}+
\frac{2\sigma}{m_g} \lvert p \rvert \delta(u)\theta(-y)(m_g y-1) ~.
\label{metfluc1}
\end{aligned}
\end{equation}
Plugging (\ref{metfluc1}) into (\ref{ansatz}) yields a flat space to the right of the particle and a curved one to the left of the particle as the particle is moving in the $+x$ direction with the speed of light. For the $\mu<0$ case, the {\it  left} and {\it right} are interchanged in the previous sentence, hence the parity-non invariance of the theory.  

Consider a massless, spinless particle traversing this geometry at an impact parameter $y=b>0$. When, (\ref{metfluc1}) is plugged in (\ref{ansatz}), there is a discontinuity in the $u$ coordinate which  can be removed by defining a new discontinuous null coordinate as
\begin{equation}
\begin{aligned}
v \equiv v_{new}-\frac{2\sigma}{m_g} \lvert p \rvert \theta(u) e^{-m_g b} ~,
\label{vmetfluc}
\end{aligned}
\end{equation}
which leads to the following time-delay for the spinless, massless particle traversing the wave
\begin{equation}
\begin{aligned}
\Delta v =-\frac{2\sigma \lvert p \rvert }{m_g} e^{-m_g b} ~.
\label{metfluc}
\end{aligned}
\end{equation}
The physical picture is as follows: as the particle traverses the  $u=0$ line, $\Delta v$ is positive (a time-delay) as long as $\sigma$ is negative for any value of the impact parameter.  Therefore unitarity and causality are not in conflict in TMG for these null geodesics.  We should note that as the mass of the graviton goes to infinity, TMG reduces to the pure Einstein gravity and there is no time-delay since there are no gravitons as we discussed before.  It is important to note that this does not say that there is no interaction between these moving particles. In fact it is well-known that even though the particles at rest do not interact in pure 3 dimensional Einstein's gravity, they do interact when one or both of them start moving. But this interaction is instantaneous \cite{ djth} and the problem of causality, if there is, is not a local one that we explore here but a global one. 

Another thing to note is to be careful about the interpretation of the results in non-Cartesian coordinates. For example, let us consider the $y < 0$ region of the shock wave 
that we discussed here. We noted that it is not in Cartesian coordinates, namely the metric function is given as $H(u,y)=  c(u) y + d(u) $, which naively yields a time-delay for any finite $b$, 
in fact an increasing time-delay when the impact parameter increases, which is simply non-physical. But, as noted in the GR part, the same situation holds in pure Einstein's gravity. Let us  repeat it: setting 
$\sigma R_{\mu \nu} =   T_{\mu \nu}$ ostensibly gives a shock wave for the pure Einstein's gravity in the form $\sigma H(u,y) \simeq -{2\lvert p \rvert}\delta(u)\theta(y) y$. But this is just flat space in another gauge (coordinates) and so there is no local gravity and no time-delay.

\subsection{Eikonal scattering in TMG}

Let us now consider obtaining the same result from the point of view of scattering amplitudes. Under the same assumptions as of section \ref{sec2}, we compute the scattering amplitude of two massless scalar particles in TMG. The scalar field coupling reads
\begin{equation}
\mathcal{S}_{TMG} = \int d^{3}x \left( \sqrt{-g} \sigma R + \frac{1}{2\mu}\epsilon^{\lambda\mu\nu} \Gamma^{\rho}_{\lambda\sigma}(\partial_{\mu} \Gamma^{\sigma}_{\rho\nu} + \dfrac{2}{3}\Gamma^{\sigma}_{\mu\alpha}\Gamma^{\alpha}_{\nu\rho}) - \frac{1}{2\alpha} k^\nu k_\nu + \frac{1}{2} g^{\mu\nu} \partial_{\mu}\phi \partial_{\nu}\phi\right) ~,
\end{equation}
where $k^{\nu}=\partial_{\mu}(\sqrt{-g}g^{\mu\nu})$. Taking into account that we deal with an eikonal scattering where the incoming momenta are
\begin{equation}
p_{1\mu} = (p_{u},\dfrac{q^2}{16p_{u}},\dfrac{q}{2}) ~, \qquad p_{2\mu} = (\dfrac{q^2}{16p_{v}},p_{v},-\dfrac{q}{2}) ~,
\end{equation}
or, in terms of Mandelstam variables, $s = -2 p_1 \cdot p_2 \simeq p_u p_v$ and $t \simeq - q^2$, $t/s \ll 1$. By using the graviton propagator in the Feynman gauge ($\alpha = 1$) (\ref{propTMG}), we can readily obtain
\begin{equation}
\begin{aligned}
\mathcal{A}_{tree}=-\sigma \frac{s^2}{t}\dfrac{1}{\left(1-\dfrac{i\sigma}{\mu}\sqrt{-t}\right)} ~.
\end{aligned}
\end{equation}
And when we calculate the phase-shift, with the prescription given in section \ref{sec2}, it gives (for $b>0$) the same result as the one obtained by the geodesic analysis.

%


\subsection{Scalar Particle in a shock wave}
The equation for a minimally coupled massless scalar field
\begin{equation}
\square\phi=0 ~,
\end{equation}
in the shock wave background reduces to
\begin{equation}
\partial_u\partial_v\phi+H(u,y)\partial^2_v\phi-\frac{1}{4}\partial^2_y\phi=0 ~,
\end{equation}
which we have not been able to solve exactly but this is not needed for our purposes. If we look at this differential equation near the wave and drop the terms that do not involve derivatives along the null directions, we end up with a solvable equation.
\begin{equation}
\partial_u\partial_v\phi+H(u,y)\partial^2_v\phi=0 ~.
\label{scalarp}
\end{equation}
After carrying out an integration in the $v$-coordinate and dropping the constant term to have zero field in the asymptotic region,  one has
\begin{equation}
\partial_u\phi+H(u,y)\partial_v\phi=0 ~.
\label{scalarp1}
\end{equation}
Now we can use the separation of variables technique to solve this equation. Assume that the  solution is in  the form $\phi(u,v,y)=U(u)V(v)Y(y)$. If we substitute this into  (\ref{scalarp1}), we get
\begin{equation}
\frac{1}{H(u,y)}\frac{U'(u)}{U(u)}+\frac{V'(v)}{V(v)}=0 ~.
\end{equation}
Let $p_v$ be the momentum of the scalar field in the $v$ direction, then
\begin{equation}
\begin{aligned}
\frac{V'(v)}{V(v)}=-\frac{1}{H(u,y)}\frac{U'(u)}{U(u)}=ip_v ~.
\end{aligned}
\end{equation}
Finally, the single mode solution reads
\begin{equation}
\phi(u,v,y)=Y(y)U(u_0)V(0)e^{ip_v \big (v-\int^u H(u',y)du'\big )} ~,
\end{equation}
from which a wave packet can be obtained by Fourier transform but this is not needed.  We assume that we know the momentum of the test particle, then we can calculate the phase it picks up when it crosses the shock wave at an impact parameter $b$ as 
\begin{equation}
\phi(0^{+},v,b)=e^{-ip_{v}\int_{0^{-}}^{0^{+}} du H(u,b)} \phi(0^{-},v,b)=e^{-ip_{v}\Delta v}  \phi(0^{-},v,b) ~,
\end{equation}
with $\Delta v$ given in (\ref{metfluc}).  Therefore, when crossing the wave, the scalar particle picks up a phase-shift akin to the Aharanov-Bohm phase. This is the same result as (\ref{metfluc}) which was computed by the light-like geodesic analysis.

\subsection{Photon in a TMG Shock Wave}

Minimally coupled photon to the shock wave gives the same result as the scalar field and the null-geodesic case that we discussed. To see the potential differences, let us now  consider a 2+1 dimensional photon non-minimally coupled to TMG given by the following action
\begin{equation}
S=-\frac{1}{4}\int d^3 x\sqrt{-g}\bigg(F_{\mu\nu} F^{\mu\nu}+\gamma R^{\mu\nu}\,_{\rho\sigma}F_{\mu\nu}F^{\rho\sigma}\bigg) ~,
\label{non_minimal_photon}
\end{equation}
where the second term, the non-minimal coupling, can be thought of as being generated after massive particles are integrated out. So this action defines an effective theory with $\gamma$  denoting a coupling constant of $L^2$ dimensions. For the shock wave background (\ref{ansatz}), the wave equation for the photon  reduces to
\begin{equation}
\nabla^\sigma F_{\rho\sigma}-\gamma R_\rho\,^{\sigma\mu\nu}\nabla_\sigma F_{\mu\nu}=0 ~, \hskip 1 cm  R_{uyuy}=-\frac{1}{2}\partial_y^2 H(u,y) ~.
\end{equation}
In components one has 
\begin{equation}
\partial_u F_{v y}+\bigg(H(u,y)+\gamma \partial_y^2 H(u,y)\bigg)\partial_v F_{v y} + \frac{1}{2} \partial_y F_{u v}=0 ~.
\end{equation}
Now, let $\epsilon_y$  represent the transverse polarization vector. The vector potential can be written as  $A_y=g(u,v)\epsilon_y$ yielding the field-strength $F_{vy}=\partial_vg(u,v)\epsilon_y$. With these, the wave equation becomes  similar to  (\ref{scalarp})
\begin{equation}
\partial_u\partial_v g(u,v)+\bigg(H(u,y)+\gamma \partial_y^2 H(u,y)\bigg)\partial^2_v g(u,v)=0 ~.
\end{equation}
Therefore, the calculation of $\Delta v$ is also similar and one arrives at the expression
\begin{equation}
\Delta v =-\frac{2\sigma \lvert p\rvert }{m_g}\big(1+\gamma  m_g^2 \big )e^{-m_g b} ~.
\end{equation}
Unlike the scalar particle or the null geodesic case,  setting $\sigma <0$, does not guarantee causality, since, depending on the sign and magnitude of  $\gamma m_g^2$, one can have a time advance  instead of a time-delay!  This is not surprising because, we considered a photon  that is not minimally coupled to gravity.  It has been been known for a long time that non-minimal coupling, which is a break-down of strong equivalence principle, could lead to superluminal motion and possibly to causality violations \cite{Drummond, Shore}. For all $\gamma >0$, there is a time-delay but if $ \gamma <0$, one must have $\gamma m_g^2 > -1$, to have a time-delay.

\subsection{ Graviton in a TMG shock wave} 
To study the gravitons in the background of the shock-wave,  let us now consider the linearization of (\ref{tmg}) (for the case of a vanishing source) about the shock-wave solution (\ref{metfluc1}) for $y>0$ then one gets the linearized field equations as
\begin{equation}
\sigma \delta G_{\mu\nu}+\frac{1}{\mu} \delta C_{\mu\nu}=0 ~.
\label{linearizedtmg}
\end{equation}
Defining the metric  functions  in the  light-cone gauge as 
\[ h_{\mu \nu}(u,v,y)= \left( \begin{array}{ccc}
g & 0 &  f \\
0 & 0 & 0 \\
 f & 0 & h\end{array} \right) ~, \]
one can compute the six equations which need to be solved consistently.  Some of these equations are complicated and we delegate them to  Appendix-B and  simply quote the solution here. To solve these equations, a close scrutiny  reveals that it would be best if one defines new functions as
\begin{equation}
\begin{aligned}
&f(u,v,y) \equiv e^{-m_g y} \int^vdv' \int^{v'} s(u,v'')\, dv'' ~, \\ \, 
&h(u,v,y) \equiv e^{-m_g y} \int^vdv' \int^{v'} r(u,v'')\, dv'' ~,\\ \, 
&g(u,v,y) \equiv e^{-m_g y}  \int^vdv' \int^{v'} p(u,v'') \, dv'' ~,\\ \, 
\end{aligned}
\end{equation}
where $s(u,v) $, $r(u,v) $ and $p(u,v) $ are functions to be determined. After plugging these into TMG field equations, the solution follows as
\begin{equation}
\begin{aligned}
&s(u,v)=e^{ip_v \Big (v-\frac{3}{2}\int^u  H(u',y) du' \Big )} ~,\\
&r(u,v) =-\frac{1}{m_g}\partial_vs(u,v) ~,\\
&p(u,v)=-\Big (\frac{ip_v H(u,y)}{2m_g}-\frac{i}{p_v}m_g \Big )s(u,v) ~.
\end{aligned}
\end{equation}
We see that $g(u,v,y) $ and $h(u,v,y) $ can be written in terms of  $f(u,v,y) $.   Again, wave-packets of gravitons, having real profiles instead of the complex ones that we have, can be constructed from this  monochromatic solution, but this is not needed as we already noted. Calculation of the time-delay follows as before and one has a shift in the graviton's phase at is crosses the shock wave as
\begin{equation}
\begin{aligned}
f(0^{+},v,b)&=e^{ \frac{-3ip_v}{2}\int_{0^{-}}^{0^{+}}  H(u',y) du' } f(0^{-},v,b) = e^{-ip_{v}\Delta v}  f(0^{-},v,b) ~.
\end{aligned}
\end{equation}
Then, $\Delta v$ yields
\begin{equation}
\Delta v =-\frac{3\sigma \lvert p\rvert }{m_g}e^{-m_g b} ~,
\end{equation}
which is positive for $\sigma <0$. It is refreshing to see that no new condition, that are not already imposed by unitarity, comes from the causality in TMG and gravitons get a time-delay.

\section{Quadratic Gravity}
The Lagrangian density of  general quadratic gravity
\begin{equation}
{\cal L}=\sqrt{-g}\bigg(\sigma R+\alpha R^2+\beta R^2_{\mu\nu}+{\cal{L}}_{\rm matter}\bigg) ~,
\label{QGaction}
\end{equation}
has the field equations
\begin{equation}
\begin{aligned}
&\sigma (R_{\mu\nu}-\frac{1}{2}g_{\mu\nu}R)+2\alpha R(R_{\mu\nu}-\frac{1}{4}g_{\mu\nu}R)+(2\alpha+\beta)(g_{\mu\nu}\square-\nabla_\mu \nabla_\nu )R\\
&\qquad +\beta\square(R_{\mu\nu}-\frac{1}{2}g_{\mu\nu}R) +2\beta(R_{\mu\sigma\nu\rho}-\frac{1}{4}g_{\mu\nu}R_{\sigma\rho})R^{\sigma\rho}=T_{\mu\nu} ~.
\label{nmgf}
\end{aligned}
\end{equation}
This theory has a massive spin-$2$ and a massive spin-$0$ graviton, with masses about the flat-space, respectively given \cite{gullucanonical} as
\begin{equation}
m^2_g=-\frac{\sigma}{\beta}~, \hskip 1cm m^2_s=\frac{\sigma}{8\alpha+3\beta} ~.
\end{equation}
Canonical analysis shows that one of these massive modes has to decouple to have a unitary theory. For the massive spin-$2$ choice the only possibility is to set  
\begin{equation}
8\alpha+3\beta=0 \label{36bis}
\end{equation} 
and choose $\sigma<0$ and $\beta>0$. This theory is called the new massive gravity (NMG) which we shall specify from now on. For the shock wave ansatz (\ref{ansatz}),  (\ref{nmgf}) reduces to
\begin{equation}
-\sigma\partial_y{^2}H(u,y)-\beta \partial_y{^4}H(u,y)=2\lvert p \rvert \delta(y)\delta(u) ~,
\label{eqnnmg}
\end{equation}
whose asymptotically flat solution is
\begin{equation}
\begin{aligned}
H(u,y) =-\frac{\sigma \lvert p \rvert\delta(u)}{ m_g}\left(e^ {-m_g\lvert y \rvert}+m_g\lvert y\rvert\right) +c_{1}y+c_{2}.
\end{aligned}
\end{equation}
Since the theory is parity invariant both the left and the right regions of the source are curved in sharp contrast to the case of TMG. By gauge-fixing we can choose the constants $c_{1}$ and $c_{2}$ in a way that the solution reads
\begin{equation}
\begin{aligned}
H(u,y) =-\frac{\sigma \lvert p \rvert\delta(u)}{ m_g} e^ {-m_g\lvert y \rvert} + 2\sigma \lvert p \rvert\delta(u) \Theta(-y)y.
\end{aligned}
\end{equation}
Without going into further detail, calculation of the time-delay yields for $b>0$
\begin{equation}
\begin{aligned}
\triangle v =-\frac{\sigma \lvert p \rvert}{m_g}e^ {-m_g\lvert b\rvert} ~,
\label{NMGtimedelay}
\end{aligned}
\end{equation}
which is positive when $\sigma$ is negative. Therefore causality and unitarity in NMG are compatible for null geodesics. 

The Eikonal scattering amplitude computation in NMG can be computed by taking into account the corresponding graviton propagator (\ref{propNMG}), which leads to
\begin{equation}
\begin{aligned}
\mathcal{A}_{tree}=-\sigma \dfrac{s^{2}}{t}\,\frac{1}{ \Big (1 - \sigma \frac{q^{2}}{m_g^2}\Big) } ~.
\end{aligned}
\end{equation}
And when calculating the phase-shift again the result is in agreement with the one obtained by the geodesic analysis.

\subsection{Photon in an NMG Shock Wave}
For the non-minimally coupled photon described by (\ref{non_minimal_photon}) coupled to the NMG shock-wave, calculation of the time-delay yields
\begin{equation}
\begin{aligned}
\Delta v &=-\frac{\sigma \lvert p\rvert }{m_g}(1+  \gamma m_g^2) e^ {-m_g\lvert b\rvert} ~.\\
\end{aligned}
\end{equation}
As long as  $\gamma m_g^2  > -1$ and $\sigma<0$,  there is a time-delay for these photons for any $b \ne 0$ impact parameter.

\subsection{Graviton in an NMG Shock Wave}
Defining the metric in the light-cone gauge as in the case of TMG and defining the following functions
\begin{equation}
\begin{aligned}
& f(u,v,y) \equiv e^{-m_g y} \int^vdv' \int^{v'} \, dv'' \int^{v''}s(u,v''') dv''' ~,\\ \, 
& h(u,v,y) \equiv e^{-m_g y} \int^vdv'\int^{v'}dv'' \int^{v''} r(u,v''') \, dv''' ~,\\ \, 
&g(u,v,y) \equiv e^{-m_g y} \int^vdv' \int^{v'}dv'' \int^{v''} p(u,v''') \, dv''' ~, \\ \, 
\end{aligned}
\end{equation}
where $s(u,v) $,$r(u,v) $ and $p(u,v) $ are the functions to be determined, one can solve the NMG equations consistently. In  Appendix-B  we give the equations for the $h =0$ case for the sake of simplicity. The solution reads as
\begin{equation}
\begin{aligned}
&r(u,v)=p(u,v)=e^{ip_v \Big (v-2\int^u  H(u',y) du' \Big )} ~,\\
&s(u,v) =-\frac{1}{m_g}\partial_vr(u,v) ~.\\
\end{aligned}
\end{equation}
We see that $r(u,v) $ and $p(u,v) $ can be written in terms of  $s(u,v) $. Then one obtains
\begin{equation}
\begin{aligned}
f(0^{+},v,b)&=e^{ -2ip_v\int_{0^{-}}^{0^{+}}  H(u',y) du' } f(0^{-},v,b) =e^{-ip_{v}\Delta v}  f(0^{-},v,b) ~.
\end{aligned}
\end{equation}
Then, calculation of $\Delta v$ yields
\begin{equation}
\Delta v =-\frac{2\sigma \lvert p\rvert }{m_g}e^{-m_g b} ~,
\end{equation}
which is positive for $\sigma <0$ and hence NMG gravitons get a time-delay. Then causality and unitarity in NMG are compatible.

\subsection{Born-Infeld Gravity}
Causal propagation in Born-Infeld type actions, be it in electrodynamics or gravity, is not automatic. Let us consider the special case of Born-Infeld extension of new massive gravity (BINMG) given by the action \cite{gullu}
\begin{equation}
I=-4 m^2\int d^3x\bigg[\sqrt{-\det(g+\frac{\sigma}{m^2}G)}- \left(1-\frac{\lambda_0}{2}\right)\sqrt{-g}\bigg] ~,
\label{BIGaction}
\end{equation}
where $G$ is matrix whose components are those of Einstein tensor $G_{\mu \nu} = R_{\mu \nu} - g_{\mu \nu} R/2$. This theory has the same spectrum as NMG around the flat and AdS backgrounds with the added property that it has a unique AdS vacuum. The field equations are cumbersome, but they are given in \cite{Gullu:2010st} and hence we shall simply use them. For $\lambda_0=0$, and for the shock-wave  ansatz the field equations, owing to the fact that $ R_{\mu\nu}R^{\mu\nu}=0,\ R_\mu^{\ \sigma}R_{\sigma\lambda}R^{\lambda\mu}=0,\ R=0$, reduce to those of NMG as
\begin{equation}
\sigma R_{\mu\nu}+\frac{1}{m^2}\square R_{\mu\nu}= \frac{1}{2}T_{\mu\nu} ~,
\label{BINMGshockeqn}
\end{equation}
which simply says that as in the case of NMG, causality and unitary are not in conflict.


\section{Anti-de Sitter space}

Let us now consider the case of negative cosmological constant. We will consider in AdS$_3$ an experiment with  the shock-waves similar to the one used in flat space. In fact, our analysis will parallel that of previous sections adapting it to the case $\Lambda <0$. Let us begin by considering the AdS$_{3}$ metric written in Poincar\'e coordinates, namely
 \begin{equation}
  ds^2_{\text{AdS}_{3}}=\dfrac{\ell^2}{y^2}(-2dudv+dy^2),
 \end{equation}
with $y\in \mathbb{R}_{\geq 0}$, $u\in \mathbb{R}$, $v\in \mathbb{R}$. The first step would be finding the metric produced by a high-energy particle located at $u=0$ and $y=y_0$. To obtain this metric we consider the Kerr-Schild ansatz 
\begin{equation} \label{ansatz1}
 ds^{2}=\dfrac{\ell^2}{y^2}\Big(-2dudv-F(u,y)du^2+dy^2\Big),
\end{equation}
and then source the field equations of the massive gravity theory with the energy-momentum tensor corresponding to such a particle. The only non-zero component of this tensor is 
\begin{equation}
T_{uu}=|p|\dfrac{\ell}{y_{0}}\delta(u)\delta(y-y_0). \label{ST}
\end{equation}

We will consider first the field equations of TMG. Later, we will consider other massive theories like NMG and MMG. 

\subsection{Topologically Massive Gravity}

The equations of motion of TMG in absence of sources take the form (\ref{tmg}); namely
\begin{equation}\label{lasTMG}
-G_{\mu\nu} + \frac{1}{\ell^2} g_{\mu\nu} + \frac{1}{\mu} C_{\mu\nu} =0 ,
\end{equation}
where $\ell = 1/\sqrt{|\Lambda |}>0$ is the AdS$_3$ radius. Notice the minus sign in front of the Einstein tensor, which is consistent with having chosen $\sigma =-1$. This makes (\ref{lasTMG}) free of ghosts around the AdS$_3$ vacuum. 

Field equations (\ref{lasTMG}) are of third order in the metric. Coupling the energy-momentum tensor (\ref{ST}) to these equations and considering the ansatz (\ref{ansatz1}) yield 
\begin{equation} \label{tmgeq}
-y\dfrac{\partial_{y}^{3}F}{2\ell \mu}-\dfrac{y^2\partial_{y}^2F-y\partial_{y}F}{2y^2}=|p|\dfrac{\ell}{y_{0}}\delta(u)\delta(y-y_{0}),
\end{equation}
which is the only non-trivial equation to solve. We know from \cite{Bending} the solutions for the homogeneous part of this equation, which would be the solutions to (\ref{tmgeq}) at both sides of the shock-wave. These solutions are
\begin{equation} \label{homo}
 F_{h}(y)=c_1\left(\dfrac{y}{\ell}\right)^{1-\ell \mu}+c_2\left(\dfrac{y}{\ell}\right)^2+c_3,
\end{equation}
where $c_1$, $c_2$ and $c_3$ are functions of $u$ but we will suppress this dependence. The  $c_2$ and $c_3$ terms can be removed by a coordinate transformation \cite{Ayon}. These correspond to the two pure-gauge modes of 3-dimensional gravity. To obtain the inhomogeneous solution to (\ref{tmgeq}), let us consider the proposal $F_{p}(y)=\theta(y-y_0)g(y)$, with $g(y)$ being of the type of (\ref{homo}), and then use matching conditions at $y_{0}$ to determine the coefficients $c_i$.  Plugging $F_{p}(y)$ into (\ref{tmgeq}) and integrating the resulting equation between a segment $y_{0}-\varepsilon$ and $y_{0}+\varepsilon$ and then taking the limit $\varepsilon \to 0$ to get rid of the delta function and its derivatives, and demanding both $F(y)$ and ${F}'(y)$ to be continuous at $y=y_0$, one finds
\begin{equation}
 {g}''(y_{0})=-2\mu\left(\dfrac{\ell}{y_0}\right)^2\delta(u)|p|,
\end{equation}
where we took $g'(y_0)={g}(y_0)=0$ for the mentioned continuity of $F(y)$ and ${F}'(y)$ --the primes denote derivatives with respect to $y$--. Then, the general solution $F_{h}(y)+F_{p}(y)$ takes the form
\begin{eqnarray}
 &F(y)= \ell^2\mu\dfrac{\delta(u)|p|}{1-(\ell \mu)^2}\left[2\left(\dfrac{y}{y_0}\right)^{1-\ell \mu}-(1-\ell \mu)\left(\dfrac{y}{y_0}\right)^2-(1+\ell \mu)\right]\theta(y-y_0)\nonumber \\[0.5em] 
 & + \ell^2\mu\dfrac{\delta(u)|p|}{1-(\ell \mu)^2}\left[2c_1\left(\dfrac{y}{y_0}\right)^{1-\ell \mu}+(1-\ell \mu)c_2\left(\dfrac{y}{y_0}\right)^2+(1+\ell \mu)c_3\right], \label{A39}
\end{eqnarray}
where $c_i$ will be determined by imposing the appropriate boundary conditions.

\subsubsection{The flat spacetime  limit}

As a consistency check of (\ref{A39}) one can verify that in the limit $\Lambda \rightarrow 0$ the result for the shock-wave profile reproduces the one for flat space found in our previous sections. This limit is also useful to gain intuition about how to set the boundary conditions. To take this limit it is convenient to define the coordinate
\begin{equation} \label{change}
 y=\ell e^{z/\ell},
\end{equation}
with which the AdS$_3$ metric takes the form
\begin{equation}
 ds^2=-2 e^{-2z/\ell} dudv+dz^2,
\end{equation}
and then one can take the limit $\ell\rightarrow\infty$ to recover the Minkowski space. In addition, we have to take into account that the profile function defined in Section (III) is related to the new profile function as
\begin{equation}
 -\dfrac{\ell^2}{y^2}F(y)=H(y).
\end{equation}

Therefore, in the limit $\ell \to \infty $ one obtain
\begin{eqnarray}
 &H(z)=\dfrac{\delta(u)|p|}{\mu}\left[2e^{-\mu(z-z_0)}-2+2\mu(z-z_0)\right]\theta(z-z_0)  \\[0.5em] \nonumber
 &  + \dfrac{\delta(u)|p|}{\mu}\left[2c_1e^{-\mu(z-z_0)}+(c_2+c_3)-\ell \mu(c_2-c_3)-2\mu c_2(z-z_0)\right], \nonumber
\end{eqnarray}
which is consistent with the result for the flat space. This provides us with a criterion to set the integration constants in (\ref{A39}). To actually have asymptotically flat space in the limit $\ell\rightarrow\infty$ we should set $c_1=0$ and $c_2=c_3$. Then, if we want to impose exactly the same boundary conditions as in our flat space analysis (asymptotically flat and Cartesian for $z>z_0$) we should also set $c_2=c_3=1$.

\subsubsection{Asymptotically AdS$_3$ boundary conditions}

Consistently, the same values for $c_1$, $c_2$ and $c_3$ are obtained by demanding asymptotically AdS$_3$ boundary conditions for finite $\ell $. In fact, one may impose the Brown-Henneaux (BH) boundary conditions \cite{BH}, demanding the metric perturbations $h_{\mu \nu }=g_{\mu \nu}-g^{\text{AdS}}_{\mu \nu}$ to be of the following orders at infinity
\begin{equation}
 h_{uu}\simeq h_{uv}\simeq h_{vv}\simeq h_{yy}\simeq \mathcal{O}(1) , \quad h_{uy}\simeq h_{vy}\simeq \mathcal{O}(y),
\end{equation}
where $\mathcal{O}(y^n)$ means that a given component decays as $y^n$ or faster close to the AdS$_3$ boundary (i.e. around $y=0$). To see this more explicitly, it is convenient to define  new coordinates as $r \equiv y^{-1}$, $t \equiv \ell^2(v+u)/2$, and $\phi \equiv \ell(v-u)/2$, for which the  components decay as 
\begin{equation}
 h_{tt}\simeq h_{t\phi }\simeq h_{\phi \phi }\simeq \mathcal{O}(1) .
\end{equation}
This means that for the type of perturbations considered in (\ref{ansatz1}) we have to impose $F(y)\sim\mathcal{O}(y^2)$. If we assume that $\mu>{1}/{\ell}$, to satisfy the BH boundary conditions we have to set $c_1=0$. To fix $c_2$ and $c_3$, we demand to have regular AdS$_3$ space deep into the bulk (i.e. at $y\rightarrow\infty$); this implies $c_2=c_3=1$. The gauge-fixed solution finally reads
\begin{eqnarray}
 &F(y)= \ell^2\mu\dfrac{\delta(u)|p|}{1-(\ell \mu)^2}\left[2\left(\dfrac{y}{y_0}\right)^{1-\ell \mu}\right]\theta(y-y_0)\\[0.5em] \nonumber
&\hskip 4.5 cm  + \ell^2\mu\dfrac{\delta(u)|p|}{1-(\ell \mu)^2}\left[(1-\ell \mu)\left(\dfrac{y}{y_0}\right)^2+(1+\ell \mu)\right]\theta(-(y-y_0)), \label{Colombo}
\end{eqnarray}
Notice that for $y<y_0$ the dependence on the $y$ coordinate is quadratic; however, this is nothing but AdS$_{3}$ written in different coordinates.

\subsubsection{Shapiro time-delay in AdS}

Let us consider the interaction of a massless particle and the shock-wave in AdS found above. The idea is to verify whether the shift in the coordinate $v$ suffered by the particle crossing the wave is positive-definite. If so, this would prevent the particle from experiencing causality problems. The equation for a massless scalar field in the shock-wave background is (after again dropping derivatives along the transverse direction)
\begin{equation}
\partial_{u}\partial_{v}\phi+F(u,y)\partial_{v}^2\phi=0.
\end{equation}
From here, we obtain the shift in the $v$ coordinate as
\begin{equation}
 \Delta v=\int_{0^{-}}^{0^{+}}du\ F(u,y) .
\end{equation}
For a particle crossing the shock-wave at $z>z_0$, we find
\begin{equation}
 \Delta v=\dfrac{2\mu|p|}{\mu^2-{1}/{\ell^2}}e^{-(\mu-1/\ell)(z-z_0)}. \label{BoundTMG}
\end{equation}
This result is sensitive to the sign of the Einstein-Hilbert term in the action --in fact, there is an implicit factor $-\sigma $ multiplying the right hand side of (\ref{BoundTMG})--. We have chosen the minus sign that renders the theory ghost free and this yields $\Delta v >0$. 

Expression (\ref{BoundTMG}) can be written as
\begin{equation}
 \Delta v= 2N \ \dfrac{|p|}{m_{g}}e^{-m_{g}(z-z_0)},
\end{equation} 
with the effective mass\footnote{It is worthwhile not to mistake this effective mass appearing in $\Delta v$ for the graviton mass of TMG around AdS$_3$, which is given by $\sqrt{\mu^2  - 1/\ell^2}$. A general analysis on the effective mass of non-linear gravitational waves in AdS$_3$ in the full theory, consisting of TMG coupled to NMG has been done in reference \cite{Bending}.} $m_{g} = \mu-1/\ell >0$ and $N=\mu/(\mu+1/\ell )>0$. This yields the result (\ref{metfluc}) in the limit $\ell \to \infty $, where $m_{g}=\mu$ and $N=1$. Therefore, the positivity of the Shapiro time-delay (\ref{BoundTMG}) expresses that unitarity and causality are compatible in TMG. 

 \subsubsection{Chiral gravity}

A particularly interesting case corresponds to the so-called chiral point $\mu\ell = 1$, where the effective mass $m_{g}$ vanishes. At this point of the parameter space, one of the central charges of the dual boundary theory (say the one of the left-moving sector, $c_L$) vanishes and this is taken as an indication that it could correspond to a chiral CFT \cite{CG}. The bulk theory also exhibits peculiar properties at the chiral point, such as the existence of new sectors of solutions. The theory defined by the TMG Lagrangian with $\mu\ell = 1$ and imposing BH boundary conditions is known as chiral gravity; it does not contain local degrees of freedom \cite{CG} and thus it is compatible with the choice $\sigma =1$. On the other hand, at $\mu\ell = 1$ the theory admits other kind of asymptotically AdS$_3$ boundary conditions \cite{GrumillerJohansson, GrumillerJohansson2}, which are weakened with respect to BH, yielding logarithmically decaying modes and do contain local degrees of freedom and thus demands again $\sigma =-1$. 

To obtain the solution at the chiral point, we have to take in (\ref{tmgeq}) the limit $\mu\ell\rightarrow1$. To do so, instead of considering the natural basis $ \{y^{1-\ell \mu}, y^2, 1\}$ for the homogeneous solution $F_h(y)$, let us take the basis $\{ (y^{1-\ell \mu}-1)/(1-\ell \mu), y^2, 1\}$. The latter is convenient as it makes explicit that in the limit $\ell \mu\rightarrow 1$ the modes of $F_h(y)$ are $\{\log(y), y^2,1\} $. If we naively take in (\ref{tmgeq}) the limit $\mu\ell \to 1$ we find
\begin{eqnarray}
 & F(y)=\dfrac{\delta(u)|p|}{\mu}\left[\log\left(\dfrac{y}{y_0}\right)-\dfrac{1}{2}\left(\left(\dfrac{y}{y_0}\right)^2-1\right)\right]\theta(y-y_0)\nonumber\\[0.5em]
 & +\dfrac{\delta(u)|p|}{\mu}\left[c_1\log\left(\dfrac{y}{y_0}\right)+c_2\left(\dfrac{y}{y_{0}}\right)^2+c_3\right]. \label{GHG}
\end{eqnarray}

As said above, at this point there are two sets of boundary conditions that we can demand: Imposing the BH boundary conditions amounts to decouple the $\sim \log (y)$ modes, that is $c_1=0$. The coefficients $c_2$ and $c_3$ are in principle undetermined because they are related to the choice of coordinates; we gauge-fix them by removing the quadratic and constant terms for $y>y_0$. We get
 \begin{equation}
  F(y)=\theta(y-y_0)\dfrac{\delta(u)|p|}{\mu}\log\left(\dfrac{y}{y_0}\right)+\theta(y_0-y)\dfrac{\delta(u)|p|}{2\mu}\left[\left(\dfrac{y}{y_0}\right)^2-1\right].
 \end{equation}

If, instead of the BH conditions, we impose in (\ref{GHG}) the weakened boundary conditions proposed by Grumiller and Johansen in \cite{GrumillerJohansson}, which in our coordinates would allow for $F(y)\sim\mathcal{O}(\log(y))$, we find 
 \begin{equation}
  F(y)=\theta(y-y_0)\dfrac{\delta(u)|p|}{\mu}\log\left(\dfrac{y}{y_0}\right)+\dfrac{\delta(u)|p|}{2\mu}\left[c_1 \log\left(\dfrac{y}{y_0}\right)\right]+\theta(y_0-y)\dfrac{\delta(u)|p|}{2\mu}\left[\left(\dfrac{y}{y_0}\right)^2-1\right] ,
 \end{equation}
where now the value of $c_1$ is not constrained. This phenomenon is the non-linear analog of the logarithmic modes of \cite{GrumillerJohansson}. The emergence of the logarithmic decaying mode is due to the finite size effects; at $\mu\ell=1$, the Compton wave length of the massive graviton equals the AdS$_3$ radius.

Also in relation to the chiral point, notice that the expression for the time delay (\ref{BoundTMG}) appears to be singular in the naive limit $\mu\ell \to 1$ which might seem puzzling as in that limit, provided one decouples the logarithmic mode, the theory is expected to be dynamically trivial; that is, one expects the chiral gravity to lose the local degrees of freedom when Brown-Henneaux boundary conditions are imposed. This seems to be in contradiction with the divergence in (\ref{BoundTMG}), as one would rather expect the time delay to vanish in that case. However the divergence is a red-herring since the expression (\ref{BoundTMG}) is not applicable at the chiral point, the reason being that some of the steps in the derivation only hold provided that $\mu\ell \neq 1$. For example, in (\ref{homo}) the gauge-fixing used to exclude the unphysical GR pure-gauge modes changes in the chiral point and no longer corresponds to setting the coefficients $c_2$ and $c_3$ of the quadratic and constant pieces to zero: As discussed above, in the limit $\mu\ell\to 1$ (or $\mu \ell \to -1$) the massive mode of TMG degenerates with the constant (respectively the quadratic) mode of GR. This means that the only dynamical mode that remains is the logarithmic one $\sim \log(y)$, while the piece that goes like $\sim y^{1\pm \mu\ell }=y^{1\pm 1}$ becomes pure gauge and can be removed by choosing $c_1=-c_3=\hat{c}/(1\pm \mu\ell)$.  Therefore, once the logarithmic mode is excluded by boundary conditions, the expression for the wave profile, $F$, reduces to that of GR, and so the rest of the computation ensues. Another step that makes (\ref{BoundTMG}) inapplicable to derive conclusions about the chiral point is the assumption $\mu\ell >1$, which yields the condition $c_1=0$ to fulfill the boundary conditions and eventually derive (\ref{Colombo}). In brief, the expression (\ref{BoundTMG}) is not applicable at the points $\mu\ell =\pm 1$, which are actually singular in many aspects. On general grounds, we could have anticipated that at the chiral point the time delay could not acquire the Yukawa dependence that (\ref{BoundTMG}) exhibits, as no massive mode survives after imposing the Brown-Henneaux AdS$_3$ boundary conditions. A similar phenomenon happens in NMG at its critical points, where the theory also exhibits degeneracy between the massive higher-derivative modes and massless GR modes. Studying the causality at the chiral points of TMG and NMG requires a separate analysis, and could be actually interesting; however, this would demand first to understand more serious issues these critical theories present when they are coupled to matter, see  \cite{Dengiz:2013hka} for example for the case of flat space limit of chiral gravity.    

\subsection{The new massive gravity}

Now, let us consider the computation in NMG. The main difference with respect to TMG is that NMG is parity-even and, consequently, contains two graviton polarizations. This is related to the fact that the NMG field equations (\ref{nmgf}) are of fourth order. Taking into account (\ref{36bis}), these can be written as follows
\begin{equation}
-G_{\mu\nu} + |\Lambda |g_{\mu\nu}+\frac{1}{2m^2}K_{\mu\nu} = 0, \label{Ufa}
\end{equation}
where $K_{\mu\nu} = 2\square R_{\mu\nu}-(1/2)\nabla_{\mu}\nabla_{\mu}R-(1/2)g_{\mu\nu}\square R+4R_{\mu\alpha\nu\beta}R^{\alpha\beta}-(3/2)RR_{\mu\nu}-g_{\mu\nu}K$; the trace $K=g^{\mu\nu}K_{\mu\nu}=R_{\mu\nu}R^{\mu\nu}-(3/8)R^2$ does not include derivatives of the curvature. This property makes NMG to be free of a scalar ghost-like degree of freedom associated to $\square R$ that other quadratic theories suffer from. This is also related to the fact that NMG coincides at the linearized with the Fierz-Pauli action for a massive spin-2 field \cite{BHT, NMG2}. It is also worth pointing out that, unlike what happens in TMG, in NMG the AdS$_3$ radius $\ell $ is not given only by $\Lambda $ but it also depends on the mass parameter $m$; see \cite{BHT, NMG2, Bending} for details.

The field equations (\ref{Ufa}) for the ansatz (\ref{ansatz1}) take the form
 \begin{equation}\label{nmgfield}
 \left[y^4\partial_{y}^{4}F+2y^3\partial_{y}^{3}F-\dfrac{(1+2\ell^2 m^2)}{2}(y^2\partial_{y}^{2}F-y\partial_{y}F)\right]\dfrac{1}{2\ell^2 m^2 y^2}=|p|\dfrac{\ell}{y_{0}}\delta(u)\delta(y-y_{0}),
 \end{equation}
whose homogeneous solutions are
\begin{equation} \label{nmghomo}
 F_{h}(y)=c_{+}\left(\dfrac{y}{\ell}\right)^{1+\beta}+c_{-}\left(\dfrac{y}{\ell}\right)^{1-\beta}+c_2\left(\dfrac{y}{\ell}\right)^2+c_3,
\end{equation}
where $c_{\pm}$ are constant coefficients and $\beta=\sqrt{1/2+\ell^2 m^2}$. Then, we proceed in a similar way as we did for TMG: First, we consider the inhomogeneous solution $F_{p}=\theta(y-y_{0})g(y)$, with $g(y)$ solving (\ref{nmgfield}), then we match at $y=y_0$ by integrating between a segment that eventually tends to zero; demanding continuity we find $g(y_0)= {g}'(y_0)= {g}''(y_0)=0$, and then we obtain
\begin{eqnarray}
{g}'''(y_0)=2m^2|p|\left(\dfrac{\ell}{y_0}\right)^{3}\delta(u).
\end{eqnarray}
The general solution to (\ref{nmgfield}) reads
\begin{eqnarray}
 &  F(y)=\theta(y-y_0)m^2|p|\delta(u)\left(\dfrac{\ell^3}{\beta^2-1}\right)\left[1-\left(\dfrac{y}{y_0}\right)^2+\dfrac{1}{\beta}\left(\dfrac{y}{y_0}\right)^{1+\beta}-\dfrac{1}{\beta}\left(\dfrac{y}{y_0}\right)^{1-\beta}\right] \nonumber \\[0.5em]
 & m^2|p|\delta(u)\left(\dfrac{\ell^3}{\beta^2-1}\right)\left[c_1-c_2\left(\dfrac{y}{y_0}\right)^2+\dfrac{c_3}{\beta}\left(\dfrac{y}{y_0}\right)^{1+\beta}-\dfrac{c_4}{\beta}\left(\dfrac{y}{y_0}\right)^{1-\beta}\right],
\end{eqnarray}
where $c_4$ is a new constant coefficient. As for TMG, these constant coefficients will be fixed by imposing suitable boundary conditions. The intuition about how to do so comes again from the flat limit $\ell\to\infty $, in which we find
\begin{eqnarray}
 &H(z)=\theta(z-z_0)|p|\delta(u)\left[2(z-z_0)-\dfrac{1}{m}\left(e^{m(z-z_0)}-e^{-m(z-z_0)}\right)\right] \nonumber \\[0.5em]
 & + |p|\delta(u)\left[-\ell(c_1-c_2)+2c_2(z-z_0)-\dfrac{1}{m}\left(c_3e^{m(z-z_0)}-c_4e^{-m(z-z_0)}\right) \right] .
\end{eqnarray}
To have asymptotically flat space we need $c_3=-1$, $c_4=0$, and $c_1=c_2$, while to have Cartesian coordinates at $z>z_0$ we need, in addition, $c_1=c_2=-1$. This finally yields
\begin{equation}
H(z)=\dfrac{|p|\delta(u)}{m}e^{-m|z-z_0|}-\theta(-(z-z_0))\dfrac{2|p|\delta(u)}{m}(z-z_0),
\end{equation}
which is the same as in our flat space computation. As in the case of TMG, the same values for $c_1$, $c_2$, $c_3$, and $c_4$ are obtained by imposing BH boundary conditions for a finite $\ell$: If we assume $m^2>1/(2\ell^2)$, then $\beta>1$, so we need $c_4=0$. If, in addition, we demand having regular AdS$_3$ deep into the bulk (i.e. $y\rightarrow\infty$) we have $c_3=-1$. The freedom in choosing $c_1$ and $c_2$ is again related to the choice of coordinates; for having AdS$_3$ in the usual coordinates at $y\rightarrow\infty$ we set $c_1=c_2=-1$. Finally, the profile for finite $\ell$ reads
\begin{equation}
 F(y)=\left(\dfrac{m^2|p|\ell^3}{\beta(1-\beta^2)}\right)\left[\theta(y-y_0)\left(\dfrac{y}{y_0}\right)^{1-\beta}
+\theta(y_0-y)\left( \left(\dfrac{y}{y_0}\right)^{1+\beta}+\beta \left( \left(\dfrac{y}{y_0}\right)^2-1 \right) \right) \right].
\end{equation}

Without going into the details in order to avoid redundancies, we can write the final result for the shift for a particle crossing the shock-wave at $z>z_0$ in NMG as
\begin{equation}\label{shiftnmg}
 \Delta v=\left(\dfrac{2m\ell^2}{2m^2\ell^2-1}\right)\dfrac{|p|}{\sqrt{1+1/(2m^2\ell^2)}}e^{\left(1/\ell-m\sqrt{1+1/(2m^2\ell^2)}\right)(z-z_0)},
\end{equation}
which can be written as
\begin{equation}\label{shiftnmg2}
 \Delta v= N \ \dfrac{|p|}{m_{g}} e^{-m_{g}(z-z_0)},
\end{equation}
with $N=m/(m+1/(2m\ell^2)+(1/\ell)\sqrt{1+1/(2m^2\ell^2)})>0$ and the effective mass given by $m_{g}= m\sqrt{1+1/(2m^2\ell^2)}-1/\ell$. This turns out to be positive-definite, and it reproduces the result of the flat space in the limit $\ell\to\infty $ (where $m_{\text{g}}=m$ and $N=1$).

\subsubsection{Critical points of NMG}

As it happens in TMG, the point of the parameter space on which $m_{\text{g}}=0$ yields vanishing central charge in the dual conformal field theory and makes the theory to acquire special properties. At this point we have $m^2\ell^2=1/2$, and this is the NMG analog to the chiral point of TMG. In the case of NMG, however, the boundary theory has no diffeomorphism anomaly, and thus one finds that both $c_R $ and $c_L$ vanish (notice that this corresponds to $\beta = 1$; see (\ref{CNMG}) below). 

In NMG, in addition, there exist another critical point, which corresponds to $m^2\ell^2= -1/2$ (that is, $\beta =0 $). The latter requires a value $m^2<0$, and in the case of asymptotically de Sitter solutions this corresponds to the partially massless point \cite{GGI}. Let us consider these two critical points separately: Let us start with $\beta =1$ and consider the point $\beta =0 $ later. For $\beta =1$ we can take the set $\{y^2\log(y),\log(y), y^2,1\} $ as the basis of $F_h(y)$. In fact, taking the limit $\beta \to 1$ in (\ref{nmgfield}), we get
\begin{eqnarray}
  & F(y)=\theta(y-y_0)|p|\delta(u)\dfrac{\ell}{4}\left(\log\left(\dfrac{y}{y_0}\right)\left[\left(\dfrac{y}{y_0}\right)^2+1\right]+\left[1-\left(\dfrac{y}{y_0}\right)^2\right]\right) \\[0.5em] \nonumber
  & +|p|\delta(u)\dfrac{\ell}{4}\left(\log\left(\dfrac{y}{y_0}\right)\left[c_1\left(\dfrac{y}{y_0}\right)^2+c_2\right]+c_3+c_4\left(\dfrac{y}{y_0}\right)^2\right) .
\end{eqnarray}

As in TMG, when $m_{\text{g}}=0$ there is more than one possible set of boundary conditions that we may consider. If we impose the BH boundary conditions, we obtain
 \begin{eqnarray}
  & F(y)=\theta(y-y_0)|p|\delta(u)\dfrac{\ell}{4}\log\left(\dfrac{y}{y_0}\right)\left[\left(\dfrac{y}{y_0}\right)^2+1\right] \\[0.5em] \nonumber
  & +|p|\delta(u)\dfrac{\ell}{4}\left(c_1\log\left(\dfrac{y}{y_0}\right)\left(\dfrac{y}{y_0}\right)^2\right)+\theta(y_0-y)|p|\delta(u)\dfrac{\ell}{4}\left(\left(\dfrac{y}{y_0}\right)^2-1\right),
 \end{eqnarray}
where the coefficient $c_1$ becomes undetermined. This indicates the presence of extra modes. The other set of boundary conditions is the one of \cite{GrumillerJohansson}, which yields
 \begin{eqnarray}
  & F(y)=\theta(y-y_0)|p|\delta(u)\dfrac{\ell}{4}\log\left(\dfrac{y}{y_0}\right)\left[\left(\dfrac{y}{y_0}\right)^2+1\right] \\[0.5em] \nonumber
  & +|p|\delta(u)\dfrac{\ell}{4}\log\left(\dfrac{y}{y_0}\right)\left[c_1\left(\dfrac{y}{y_0}\right)^2+c_2\right]+\theta(y_0-y)|p|\delta(u)\dfrac{\ell}{4}\left(\left(\dfrac{y}{y_0}\right)^2-1\right),
 \end{eqnarray}
and includes the additional logarithmic modes with coefficient $c_2$.

In the critical point $\beta = 0$, on the other hand, we find for $F_h(y)$ the modes $\{y\log(y), y,  y^2, 1\}$. In this case, the solution (\ref{nmgfield}) in the limit $\beta \to 0$ yields
\begin{eqnarray}
   & F(y)=\theta(y-y_0)|p|\delta(u)\dfrac{\ell}{2}\left(2\dfrac{y}{y_0}\log\left(\dfrac{y}{y_0}\right)+\left[1-\left(\dfrac{y}{y_0}\right)^2\right]\right) \nonumber \\[0.5em]
   & +|p|\delta(u)\dfrac{\ell}{2}\left(2c_1\left(\dfrac{y}{y_0}\right)\log\left(\dfrac{y}{y_0}\right)+c_2\left(\dfrac{y}{y_0}\right)+c_3+c_4\left(\dfrac{y}{y_0}\right)^2\right),
\end{eqnarray}
which, apart from the logarithmic modes $\sim y\log (y)$ also includes the linear mode $\sim y$ which is characteristic of conformal gravity \cite{conformal} --even though the higher-curvature terms of NMG are not conformally invariant, but conformally covariant--. This linear mode is responsible of the existence of hairy black holes in NMG around (A)dS spaces \cite{NMG2, OTT}; see also the discussion in \cite{GGI}.

\subsection{Unitarity in the boundary CFT$_2$}

Now, having shown the compatibility between unitarity and causality in the bulk, let us study the necessary conditions for unitary in the dual CFT.

\subsubsection{The bulk/boundary unitarity clash}

As already mentioned, both TMG and NMG suffer from the so-called bulk/boundary unitarity clash. That is, the conflict between the value of the coupling constants that make the bulk theory unitary and those that make the boundary theory unitary. More precisely, in the case of TMG the dual CFT$_2$ has left- and right-moving central charges given by 
\begin{equation}
c_L = \frac{3\ell }{2G} \Big( \sigma -\frac{1}{\ell \mu }\Big) \ , \ \ \ c_R = \frac{3\ell }{2G} \Big( \sigma +\frac{1}{\ell \mu }\Big),
\end{equation}
which are negative for the choice $\sigma =-1$. Reciprocally, demanding $c_L\geq 0 \leq c_R$ leads to choose the sign of the Einstein-Hilbert action with a ghost. The same occurs with NMG, which yields a dual CFT$_2$ with central charges
\begin{equation} \label{CNMG}
c_L = c_R = \frac{3\ell }{2G} \Big( \sigma - \frac{1}{2\ell^2m^2 }\Big),
\end{equation}
which also require $\sigma = 1$ to be positive. This implies that, for these theories, either the bulk theory is not unitary or the boundary CFT$_2$ is not unitary. Asking the boundary CFT$_2$ to be unitary, and therefore $c_L >0<c_R$, the mass spectrum of the BTZ black holes turns out to be positive too, yielding $L_0 \geq 0 \leq \bar{L}_0$, and this is one of the reasons why the microscopic derivation of the black hole entropy in terms of the Cardy formula works. This is also why in \cite{CG} a proposal was made to decouple the local degrees of freedom in the bulk while still keeping $c_L = 0 \leq c_R$ (namely, considering $\mu \ell =1$ with $\sigma = 1$). However, the conflict between bulk and boundary unitarity remains for general values of $\mu \ell$. 

In the last years, different theories in 3 dimensions have been proposed as proposals to solve the bulk/boundary unitarity clash. For instance, in reference \cite{ZDG} a bi-gravity theory called zwei-dreibein gravity (ZDG) was proposed, which can be regarded as a generalization of NMG. More recently, in reference \cite{berk}, an extension of TMG called minimal massive gravity (MMG) has been proposed as a model that would eventually\footnote{However, it has been observed in \cite{Nuevo} that, as it happens with the chiral limit of TMG, MMG in the metric formulation in principle contains logarithmic modes and that, if the solution for the bulk/boundary unitarity clash is met, this should also involve a special choice of boundary conditions or there must be a linearization instability of AdS vacuum in the theory.} yield a ghost-free theory about AdS$_3$ while, at the same time, yield positive values for the central charges if the dual CFT$_2$. The terms that MMG add to TMG do not contribute in asymptotically flat space for the specific solutions we have considered here. In contrast, they do contribute in the case of AdS$_3$. Therefore, it is worthy considering these terms here. We will see below that when such MMG terms are considered, the result for the Shapiro time-delay is also positive, even in the window of parameter space in which the central charges of the dual CFT$_2$ are positive.

\subsubsection{Minimal massive gravity}

The field equations of MMG are
\begin{equation}\label{eqMMG}
 -G_{\mu\nu}+ |\Lambda  | g_{\mu\nu}+\dfrac{1}{\mu}C_{\mu\nu}+\dfrac{\gamma}{\mu^2}J_{\mu\nu}=0,
\end{equation}
with $J_{\mu\nu}=R_{\mu}^{\rho}R_{\rho\nu}-({3}/{4})RR_{\mu\nu}-({1}/{2})g_{\mu\nu}(R^{\mu\nu}R_{\mu\nu}-({5}/{8})R^2)$. These correspond to adding to the field equations of TMG a second order tensor $J_{\mu \nu }$. Lovelock theorem forbids this tensor to come from a variational principle in the second order formalism. In fact, one can verify that $J_{\mu\nu}$ is not identically conserved, although it is conserved on-shell \cite{berk}. 

Considering in MMG the ansatz (\ref{ansatz1}), yields
\begin{equation}
  ds^{2}=\dfrac{\ell^{2}}{y^{2}}(-2dudv-F(u,y)du^2+dy^2),
\end{equation}
the equation of motion for a shock-wave in this theory is \cite{Yerko}
\begin{equation}
 \dfrac{1}{4\ell^{4}\mu^{3} y}\left(-(\gamma\mu-2\ell^{2}\mu^{3})\ell^{2}\dfrac{\partial F}{\partial y}+(\gamma\mu-2\ell^{2}\mu^{3})\ell^{2} y\dfrac{\partial^{2} F}{\partial y^{2}}-2\ell^{3}\mu^{2}y^{2}\dfrac{\partial^{3} F}{\partial y^{3}}\right)=|p|\dfrac{\ell}{y_{0}}\delta(u)\delta(y-y_0),
\end{equation}
whose homogeneous solutions have been found in \cite{Yerko}. Following the same procedure as in for the case of TMG, we find the Shapiro time-delay in MMG, which reads
\begin{equation}
 \Delta v=N \ \dfrac{|p|}{m_{g}}e^{-m_{g}(z-z_0)}, \label{BoundMMG}
\end{equation}
with $m_{g}=\mu-1/\ell -\gamma /(2\mu\ell^2)$ and $N= \mu/(\mu+1/\ell -\gamma/(2\mu\ell^2))$. In reference \cite{berk} it has been shown how, for the choices of parameters that yield (\ref{eqMMG}), a windows exists for the value of $\gamma $ such that, even if $\sigma =-1$, the central charges of the boundary CFT$_2$ result positive. It is easy to verify that within such window the value (\ref{BoundMMG}) turns out to be positive-definite. This shows the compatibility between bulk causality and the necessary conditions $c_L > 0 < c_R$ for the unitarity in the dual CFT$_2$. The problem with unitarity still remains due to the logarithmic modes discussed in \cite{Nuevo}. Whether or not a consistent way of decoupling such modes exists deserves further analysis. 

Let us add that the same kind of computation can be done for ZDG theory, whose homogeneous wave solutions in AdS$_3$ $F_h$ are also known explicitly \cite{Andres}.


\section{Conclusions}

We have studied the issue of local causality in $2+1$ dimensional topologically massive gravity and the new massive gravity. We have shown that unlike the quadratic and cubic theories in dimensions $n \ge 4$, causality and unitarity are not in contradiction in 3 dimensions. Namely, as long as the  sign of the Newton's constant is chosen to be the opposite to the one considered in the higher-dimensional case, TMG and NMG in asymptotically flat spacetime turn out to be causal and unitary. We have also investigated the Born-Infeld extensions of NMG, which have also been shown to be causal and unitary. We have also performed the analysis in these theories for the asymptotically Anti-de Sitter (AdS) spacetime. Again, local causality and unitarity were found to be consistent. The notion of local causality we considered here is the positivity of the Shapiro time-delay for null geodesics and minimally-coupled fields, while with unitary we mean the absence of ghost and tachyon excitations.


The observation of \cite{Camanho:2014apa} that Einstein-Gauss-Bonnet and (Riemann)$^3$ theories are not causal and can only be made causal by relying on the existence of a UV completion of the theory such as string theory, naturally give raise the question as to whether 3-dimensional (massive) gravity fits into this picture where no higher spin states are needed to ensure local causality. While a full understanding of this might require further investigations, we have some remarks to make: First of all, to the best of our knowledge, none of the theories considered here come from string theory compactifications. At least, no derivation of them without introducing extra fields is known. Some of them, however, involve Chern-Simons Lagrangians, thereby they are likely to be well-posed at quantum level. This even permits one to write down theories for a finite number of higher-spin fields (see \cite{Ammon:2012wc}, and references therein), something that is not possible in four or higher dimensions. All in all, we can interpret our results as further evidence suggesting that gravity in $3$ dimensions might exist without reference to string theory, as an effective theory not hindered by causality violation.

\section{\label{ackno} Acknowledgments}

We want to thank Xi\'an Camanho for a critical reading of the manuscript and sharing with us the contents of \cite{Xupcoming}. J.D.E. wishes to thank the Department of Physics of the University of Buenos Aires for kind hospitality while this work was being done. The work of J.D.E. is supported by MINECO (FPA2014-52218), Xunta de Galicia (GRC2013-024), the Spanish Consolider-Ingenio 2010 Programme CPAN (CSD2007-00042), FEDER and EPLANET. C.G. thanks the Solvay Institutes for the hospitality during her stay at Universit\'e Libre de Bruxelles, Belgium. The work od G.G. and M.L. is supported by CONICET (PIP 0595/13). The work of B.T. is supported by the TUBITAK Grant No.113F155. E.K. is supported by the TUBITAK 2214-A Scholarship.

\section{Appendices}

\subsection{ Shock wave geometry}

The signs play a crucial role therefore let us fix them explicitly here: the signature of the metric is $ (-,+,+)$, the Riemann tensor reads as $R^\mu\,_{\nu\alpha\beta}=\partial_{\alpha}\Gamma^\mu_{\nu\beta}-...$.

The shock wave metric describes the spacetime generated by a point particle moving at the speed of light.  Let $(t,x,y)$ be the coordinates in the Minkowski space. Defining the null coordinates as  $u=t-x$ and $v=t+x$ and considering a massless point particle moving in the  $+x$ direction with 3 momentum as $p^\mu=\lvert p\rvert(\delta^\mu_0+\delta^\mu_x) $ and with the energy momentum tensor as  $T_{uu}=\lvert p\rvert \delta(y)\delta(u)$, the ansatz for the metric created by this particle  in local coordinates read
\begin{equation}
ds^2=-du dv+H(u,y) du^2+dy^2.
\label{shockwavemetric}
\end{equation}
To simplify the relevant computations \cite{gurses_sisman}, let us write the metric in the Kerr-Schild form as  $g_{\mu \nu}=\eta_{\mu \nu}+H(u,y) \lambda_\mu \lambda_\nu$ with the $\lambda_\mu$ vector satisfying the following properties
\begin{equation}
\begin{aligned}
\lambda^{\mu}\lambda_\mu=0,\hskip .6 cm\nabla_{\mu}\lambda_\nu=0,\hskip .6 cm
\lambda^\mu\partial_\mu H(u,y)=0.
\label{plane1}
\end{aligned}
\end{equation}
In the null coordinates, non-vanishing components of $\eta_{\mu \nu}$ are  $\eta_{u v} = - \frac{1}{2}$ and  $\eta_{ yy} = 1$ and one also has  $\det{g} = \det{\eta} = -\frac{1}{4}$.
The Christoffel symbols can be found as 
\begin{equation}
2\Gamma^\sigma_{\mu \nu}= \lambda^\sigma \lambda_\mu \partial_\nu H + \lambda^\sigma \lambda_\nu \partial_\mu H - \lambda_\mu \lambda_\nu \eta^{\sigma \beta} \partial_\beta H, 
\end{equation}
whose non-vanishing components are 
\begin{equation}
 \Gamma^y_{uu}=-\frac{1}{2}\partial_y H(u,y),\hskip .6 cm \Gamma^v_{uu}=-\partial_u H(u,y),\hskip .6 cm \Gamma^v_{uy}=-\partial_y H(u,y).
\end{equation}
Observe that one has vanishing contractions $\lambda_\sigma\Gamma^\sigma_{\mu \nu}=0$, $\lambda^\mu\Gamma^\sigma_{\mu \nu}=0$ and the Riemann tensor is also linear in the derivative of the metric function $H$ as no contribution comes from the products of the connections
\begin{equation}
2 R_{\mu\alpha\nu\beta}=\lambda_{\mu}\lambda_{\beta}\partial_{\alpha}\partial_{\nu}H+\lambda_{\alpha}\lambda_{\nu}\partial_{\mu}\partial_{\beta}H
-\lambda_{\mu}\lambda_{\nu}\partial_{\alpha}\partial_{\beta}H-\lambda_{\alpha}\lambda_{\beta}\partial_{\mu}\partial_{\nu}H.\label{Riemann_pp-wave}
\end{equation}
From this follow the  Ricci tensor  and the "Box"  of the Ricci tensor (which are relevant in the NMG and the Born-Infeld gravity cases discussed in the text) as 
\begin{equation}
R_{\mu\nu}=-\frac{1}{2}\lambda_{\mu}\lambda_{\nu}\partial_y^2 H(u,y),\hskip .6 cm  \Box R_{\mu \nu} = -\frac{1}{2}\lambda_{\mu}\lambda_{\nu}\partial_y^4 H(u,y).
\label{ricci_shock}
\end{equation}
Finally,  for the TMG case, we need the Cotton tensor which is defined as
\begin{equation}
  C_{\mu\nu} = \eta_\mu{^{\alpha\beta}} \nabla_\alpha \bigg (R_{\nu\beta}- \frac{1}{4} g_{\nu\beta} R \bigg ),
\end{equation}
with $\eta^{\mu\alpha\beta}$ being the completely antisymmetric tensor normalized as \footnote{This corresponds to the sign choice $\epsilon^{t x y} =- 1$.} $ \eta^{ u v y} =  2$.  It follows that for the shock wave, one has 
\begin{equation}
C_{\mu\nu}=\frac{1}{2}\lambda_{\mu}\lambda_{\nu}\partial_y^3 H(u,y).
\label{cotton_shock}
\end{equation}
One can use these tensors to find the shock wave solutions in various gravity theories, which we have done in the relevant sections above.  Let us  also compute the spin-2 perturbations about a given shock wave background below as they are relevant to the gravitons scattering through the shock wave. 

\subsection{Perturbations about the shock wave}
In principle, one can work out the most general perturbation without choosing a gauge, but the ensuing computations are unnecessarily  cumbersome in theories beyond GR, so a proper choice of gauge that keeps all the physical degrees of freedom is important.  Defining the perturbation as $ h_{\mu \nu} \equiv \delta g_{\mu \nu}$, the light-cone gauge seems to be the best choice for our purposes. So we set 
\begin{equation}
 h_{v\mu}=0,
\end{equation}
or more covariantly we have  $\lambda^{\mu}h_{\mu\nu}=0$ and the following equations hold in this gauge
\begin{equation}
 \Gamma^\sigma_{\mu \nu}h^\mu\,_{\sigma}=0,\hskip .6 cm \Gamma^\sigma_{\mu \nu}h_{\sigma\alpha}=-\frac{1}{2}\lambda_{\mu}\lambda_{\nu}h_{y\alpha}\partial_yH
 ,\hskip .6 cm \Gamma^\sigma_{\mu \nu}h^\mu\,_{\alpha}=\frac{1}{2}\lambda^{\sigma}\lambda_{\nu}h^y\,_{\alpha}\partial_yH.
\end{equation}
The linearized connections can be calculated as 
\begin{equation}
\delta \Gamma^\sigma_{\mu \nu} = \frac{1}{2}\eta^{\sigma \alpha}\Big ( \partial_\mu h_{ \nu \alpha} + \partial_\nu h_{ \mu \alpha}  - \partial_\alpha h_{ \mu \nu}+  \lambda_\mu \lambda_\nu  h_{\alpha y} \partial_y H \Big ) - H \lambda^\sigma \partial_v h_{ \mu \nu},
\end{equation}
or more explicitly in components, one has
\begin{equation}
\begin{aligned}
&\delta \Gamma^u_{\mu \nu} = \partial_v h_{ \mu \nu},\\
&\delta \Gamma^v_{\mu \nu} = -\partial_\mu h_{\nu u}-\partial_\nu h_{\mu u}+\partial_uh_{\mu \nu}-\lambda_\mu \lambda_\nu h_{yu}\partial_y H+2H\partial_v h_{\mu\nu}, \\
&\delta \Gamma^y_{\mu \nu} =  \frac{1}{2} \Big (\partial_\mu h_{\nu y}+\partial_\nu h_{\mu y}-\partial_yh_{\mu \nu}+\lambda_\mu \lambda_\nu h_{yy}\partial_y H \Big).
\end{aligned}
\end{equation}
The linearized Ricci tensor 
\begin{equation}
\begin{aligned}
 \delta R_{\mu\nu}
 =\frac{1}{2} \Big (\nabla_\sigma\nabla_\mu h^\sigma\,_{\nu}+\nabla_\sigma\nabla_\nu h^\sigma\,_{\mu}-\square h_{\mu\nu}-\nabla_\mu\nabla_\nu h \Big ),
\end{aligned}
\end{equation}
boils down to the following form in the light-cone gauge
\begin{equation}
 \begin{aligned}
 2\delta R_{\mu\nu}&=2\partial_{(\mu}\partial_\sigma h^\sigma\,_{\nu)}+\lambda_\mu \lambda_\nu\partial_y H\partial_\sigma h^\sigma\,_{y}
 +h\lambda_\mu \lambda_\nu\partial_y^2 H - g^{ \alpha \beta} \partial_\alpha \partial_\beta h_{\mu \nu}\\
& +4\partial_y H\partial_v\lambda_{(\mu} h_{\nu)y}-\partial_\mu\partial_\nu h+\Gamma^\sigma_{\mu \nu}\partial_\sigma h,
\end{aligned}
\end{equation}
where we used the round brackets to denote symmetrization with a factor of $1/2$.  The linearized  scalar curvature reads
\begin{equation}
\delta R=\partial_{\mu}\partial_\sigma h^{\sigma\mu}-\square h.
\end{equation}
Computation of the linearization of the Cotton tensor is somewhat long, we use the form given in  \cite{Kilicarslan:2015cla} valid for an arbitrary background as 
\begin{equation}
\begin{aligned}
 2\delta C^{\mu\nu}=&-\frac{h}{2}\,C^{\mu\nu}+\eta^{\mu \rho \sigma}\,\nabla_{\rho} \delta G^\nu{_\sigma}+\eta^{\mu \rho \sigma}\,\delta \Gamma^\nu_{\rho \alpha}G^{\alpha}{_{\sigma}}
 +\mu \leftrightarrow \nu\\
 =&-\frac{3h}{2}\,{C^{\mu\nu}}-\frac{1}{2}\eta^{\mu \rho \sigma}\,\square \nabla_{\rho }h^{\nu}{_{\sigma}}
 +\frac{1}{2}\eta^{\mu \rho \sigma}\,\nabla^{\nu}\nabla_{\lambda}\nabla_{\rho }h_{\sigma}{^{\lambda}}
+ \frac{3}{2}\eta^{\mu \rho \sigma}\, \nabla_{\rho }({\cal S}^{\lambda\nu} h_{\lambda\sigma})
 +\frac{1}{6} \eta^{\mu \rho \sigma}\, R \nabla_\rho  h^\nu{_\sigma}\\
 &-\frac{1}{2}\eta^{\mu \rho \sigma}\,{\cal S}^\nu{_\sigma} \nabla_{\rho }h -\frac{1}{2}\eta^{\mu \rho \sigma}\, h^\lambda{_\sigma} \nabla_\lambda {\cal S}^\nu{_\rho }+\eta^{\mu \rho \sigma}\,{\cal S}_{\lambda\rho }\nabla^\nu h^\lambda{_\sigma}
 +\eta^{\mu \rho \sigma}\,{\cal S}_\sigma{^\lambda} \nabla_{\lambda}h^{\nu}{_{\rho }}+\mu \leftrightarrow \nu,
 \label{cottlin}
 \end{aligned}
 \end{equation}
where ${\cal S}_{\mu\nu} = R_{\mu\nu}  - \frac{1}{3}g_{\mu\nu}R$. 
For (\ref{shockwavemetric}) and in the light-cone gauge, (\ref{cottlin}) reduces to
 \begin{equation}
\begin{aligned}
 2\delta C^{\mu\nu}=&-\frac{3h}{4}\,\lambda^\mu\lambda^\nu \partial_y^3H-\frac{1}{2}\eta^{\mu \rho \sigma}\,\square \nabla_{\rho }h^{\nu}{_{\sigma}}
 +\frac{1}{2}\eta^{\mu \rho \sigma}\,\nabla^{\nu}\nabla_{\lambda}\nabla_{\rho }h_{\sigma}{^{\lambda}}\\
&-\frac{1}{2}\eta^{\mu \rho u}\,\delta^\nu_v  \partial_y^2H \nabla_{\rho }h -\frac{1}{2}\eta^{\mu u y}\, h^\lambda{_y}\delta^\nu_v \nabla_\lambda  \partial_y^2H
+\eta^{\mu y u}\,\partial_y^2H\nabla_v h^\nu{_y}+\mu \leftrightarrow \nu,
 \end{aligned}
 \end{equation}
which could still be simplified further, but this form is all we need to carry out our computations.

In the light-cone gauge, $ h_{\mu  v } =0  $, and with the definitions  $g \equiv h_{u u}$, $ f \equiv h_{u y} $, $ h  \equiv h_{y y} $, where all functions depend on all coordinates, we list the explicit forms of the linearized forms of various components about the shock-wave background. We have made use of these results while studying the graviton propagation in the relevant theories in the text.  

Components of the  linearized Ricci tensor are
\begin{equation}
\begin{aligned}
&\delta R_{uu}= (\partial_u\partial_y+ \partial_yH\partial_v)f+(2 H
 \partial_v^2 -\frac{1}{2}\partial_y^2)g
 +\frac{1}{2}(\partial_y^2 H +\frac{1}{2}\partial_yH \partial_y-\partial_uH \partial_v-\partial_u^2)h,\\
 &\delta R_{uv}= \frac{1}{2}\partial_v\partial_yf-\partial_v^2g-\frac{1}{2}\partial_u\partial_vh,\\
 &\delta R_{uy}=(2H \partial_v^2+\partial_u\partial_v)f-\partial_v\partial_yg+\frac{1}{2} \partial_yH \partial_vh,\\
 &\delta R_{vv}= -\frac{1}{2}\partial_v^2h,\\
 &\delta R_{vy}= -\partial_v^2f,\\
 &\delta R_{yy}=-2\partial_v\partial_yf+2(H \partial_v^2+\partial_u\partial_v)h.
 \end{aligned}
\end{equation}
The linearized curvature scalar reads
\begin{equation}
\delta R = 4 \left( - \partial_v \partial_y  f + \partial_v^2 g  + H \partial_v^2 h + \partial_u \partial_v h \right) ~.
\end{equation}
Components of the  linearized Cotton tensor are
\begin{equation}
\begin{aligned}
&\delta C_{uu}=\frac{1}{4} \bigg((-10\partial_yH \partial_v\partial_y+4\partial_u H \partial_v^2+16 H^2
 \partial_v^3-4H\partial_v\partial_y^2+16H \partial_v^2\partial_u+4\partial_v\partial_u^2-4 \partial_u\partial_y^2-6\partial_y^2H \partial_v)f\\&
 +(-4\partial_yH \partial_v^2-8H \partial_v^2\partial_y+2\partial_y^3g-4\partial_v\partial_u\partial_yg)g
+(\partial_yH (12 H \partial_v^2-\partial_y^2 +8  \partial_v\partial_u)+2\partial_u H \partial_v\partial_y\\&+4 \partial_u\partial_yH \partial_v +4H\partial_v\partial_u\partial_y
  +2\partial_u^2\partial_y-3\partial_y^2H \partial_y -3\partial_y^3H )h\bigg),\\  
 &\delta C_{uv}=\frac{1}{2}\partial_v^2\bigg(-(4H \partial_v+2 \partial_u)f+\partial_y g+(-\partial_yH +H\partial_y)h
 \bigg), \\
 &\delta C_{uy}=(-2 H\partial_v^2\partial_y -2 \partial_v\partial_u\partial_y-3\partial_yH 
\partial_v^2) f+(-2 H \partial_v^3 + \partial_v\partial_y^2 - \partial_v^2\partial_u )g+(\partial_uH \partial_v^2+ \partial_v\partial_u^2\\&
-\frac{1}{2}\partial_yH \partial_v\partial_y- \partial_y^2H \partial_v+H (2 H \partial_v^3 +3
\ \partial_v^2\partial_u)) h,\\
 &\delta C_{vv}=\partial_v^2(\partial_vf-\frac{1}{2} \partial_y h),\\
 &\delta C_{vy}=\partial_v^2 \left(\partial_v g-(H\partial_v+\partial_u) h\right),\\
 &\delta C_{yy}=\partial_v^2 \left(-4(H \partial_v+\partial_u)f+2\partial_y g-2h\partial_yH\right).
 \end{aligned}
\end{equation}
With these results, each component of the TMG equations can be computed. But since two components of the equations are  somewhat complicated, we shall  simplify them by using the fact that away from $y = 0$ we have $\partial_y  H = - m_g H$ 

\begin{equation}
\begin{aligned}
&\partial_v^2\Big(2\partial_v f+( m_g - \partial_y)h\Big) =0 \hskip  1 cm vv-\mbox{component}
\end{aligned}
\end{equation}
\begin{equation}
\begin{aligned}
& \partial_v\Big((-4 H\partial_v^2 +m_g \partial_y-2\partial_v\partial_u)f+\partial_v\partial_y g -( m_g  H
  \partial_v -H\partial_v \partial_y +m_g \partial_u )h\Big)= 0\\& \hskip  10 cm vu-\mbox{component}
\end{aligned}
\end{equation}
\begin{equation}
\begin{aligned}
\partial_v^2\Big(-m_g f-\partial_v g+(H \partial_v+\partial_u )h\big)  = 0 \hskip  1 cm vy-\mbox{component}
\end{aligned}
\end{equation}
\begin{equation}
\begin{aligned}
&\partial_v^2\Big(2( H\partial_v +\partial_u)f-(m_g  
 +\partial_y) g- m_g H h\Big)=0  \hskip  1 cm yy-\mbox{component}
\end{aligned}
\end{equation}

\begin{equation}
\begin{aligned}
& (m_g H \partial_v^2-2H\partial_v^2\partial_y-m_g  \partial_v\partial_u-2 \partial_v\partial_u\partial_y)f+(-2 H \partial_v^3 +m_g \partial_v\partial_y +\partial_v\partial_y^2  -\partial_v^2\partial_u )g\\
&+(2 H^2 \partial_v^3 +2H \partial_u \partial_v^2 
  + \frac{H}{2}m_g \partial_v \partial_y -\frac{H}{2}m_g^2  \partial_v+\partial_v \partial_u^2 )h
  =0  \hskip  1 cm uy-\mbox{component}
\end{aligned}
\end{equation}

\begin{equation}
\begin{aligned}
 & (-2m_g^2H\partial_v+2m_g H\partial_v\partial_y+16 H^2\partial_v^3 
 -4 H\partial_v\partial_y^2+12 H\partial_v^2\partial_u-4 m_g \partial_u\partial_y
-4\partial_u\partial_y^2+4\partial_v\partial_u^2)f\\&+ (-2m_g^2 H  \partial_y-4 m_g H^2
 \partial_v^2 + m_g H\partial_y^2 + m_g^3 H+2m_g H \partial_u\partial_v  
+2 H \partial_u\partial_v\partial_y+2m_g  \partial_u^2 +2\partial_u^2\partial_y)h\\&+(-8 H
 \partial_v^2\partial_y +2 m_g \partial_y^2 +2\partial_y^3 -4 \partial_v\partial_u\partial_y +4 m_g H \partial_v^2)g =0  \hskip  1 cm uu-\mbox{component}
\end{aligned}
\end{equation}

Consider the linearized field equations of NMG about the shock-wave background in the axial-like gauge. For this computation, we shall consider a further simplification within the light-cone gauge and assume that the perturbation is also traceless, namely $ h=0$, otherwise the linearized equations are cumbersome. Then we start with   
\[ h_{\mu \nu}(u,v,y)= \left( \begin{array}{ccc}
g & 0 &  f \\
0 & 0 & 0 \\
 f & 0 & 0\end{array} \right).
\]
With this definition, each component of the NMG equations can be computed. But since two components of the equations are  somewhat complicated, we shall  simplify them again by using the fact that we have $\partial_y  H = - m_g H$ away from the source.

\begin{equation}
\begin{aligned}
\partial_v^4g-\partial_v^3\partial_yf =0 \hskip  1 cm vv-\mbox{component}
\end{aligned}
\end{equation}
\begin{equation}
\begin{aligned}
& (-4 m_g H\partial_v^3 +m_g^2 \partial_v\partial_y+2\partial_v^2\partial_u\partial_y)f+( 
-\partial_v^2\partial_y^2+2\partial_v^3\partial_u+4H\partial_v^4)g = 0 \hskip  1 cm vu-\mbox{component}
\end{aligned}
\end{equation}
\begin{equation}
\begin{aligned}
(-4 H\partial_v^4 -4 \partial_v^3\partial_u-m^2\partial_v^2)f+\partial_v^3\partial_y g= 0 \hskip  1 cm vy-\mbox{component}
\end{aligned}
\end{equation}
\begin{equation}
\begin{aligned}
(-4 m_g H\partial_v^3 +2H \partial_v^3\partial_y+2\partial_v^2\partial_u\partial_y)f+( 
-\partial_v^2\partial_y^2+2\partial_v^3\partial_u+2H\partial_v^4+m_g^2\partial_v^2)g = 0  \hskip  .8 cm yy-\mbox{component}
\end{aligned}
\end{equation}

\begin{equation}
\begin{aligned}
& (m_g^2\partial_v\partial_u- 2\partial_v\partial_u\partial_y^2+4 \partial_v^2\partial_u^2-4 H\partial_u\partial_v^3+5 m_g H\partial_v^2\partial_y-2 m_g^2 H\partial_v^2
-2 H\partial_v^2\partial_y^2+12H\partial_v^3\partial_u\\&+8H^2 \partial_v^4)f
+(-m_g^2\partial_v\partial_y + \partial_v \partial_y^3 
 -3 \partial_v^2 \partial_u\partial_y +3m_g H  \partial_v^3-4H\partial_v^3 \partial_y )g
  =0  \hskip  1 cm uy-\mbox{component}
\end{aligned}
\end{equation}

\begin{equation}
\begin{aligned}
   &(-8mH\partial_u\partial_v^2-2\partial_u\partial_y^3+6 \partial_v\partial_u^2\partial_y
 +2m^3 H \partial_v-7\partial_y H \partial_v\partial_y^2
 +24 H \partial_yH \partial_v^3-2H \partial_v\partial_y^3\\&+10H \partial_v^2\partial_u\partial_y
 +8H^2 \partial_v^3\partial_y+2 m^2 \partial_u\partial_y-5m^2H \partial_v\partial_y)f +(-4
\partial_v\partial_u\partial_y^2+2\partial_v^2\partial_u^2-7 \partial_yH \partial_v^2\partial_y\\&-6 H \partial_v^2\partial_y^2+6 H \partial_v^3\partial_u+8 H^2
  \partial_v^4 -m^2 \partial_y^2+\partial_y^4-3\partial_y^2H \partial_v^2)g
   =0  \hskip  1 cm uu-\mbox{component}
\end{aligned}
\end{equation}

\subsection{Scattering amplitudes in Massive Gravity}

In order to compute eikonal scattering amplitudes in massive gravities, it is fairly convenient to introduce a set of orthogonal projection operators constructed from the transverse and longitudinal projectors \cite{Barnes,Rivers}
$$
\theta_{\mu\nu}=\eta_{\mu\nu}-\dfrac{\partial_{\mu}\partial_{\nu}}{\Box} ~, \qquad \omega_{\mu\nu}=\partial_{\mu}\partial_{\nu} ~.
$$
These are six operators in the space of symmetric tensor fields,
\begin{eqnarray}
P^{(2)}_{\mu\nu,\rho\sigma} &=& \dfrac{1}{2}(\theta_{\mu\rho}\theta_{\nu\sigma}+\theta_{\mu\sigma}\theta_{\mu\rho}-\theta_{\mu\nu}\theta_{\rho\sigma}) ~, \qquad P^{(0,s)}_{\mu\nu,\rho\sigma} = \dfrac{1}{2}\theta_{\mu\nu}\theta_{\rho\sigma} ~, \nonumber \\ [0.5em] 
P^{(1)}_{\mu\nu,\rho\sigma} &=& \dfrac{1}{2}(\theta_{\mu\rho}\omega_{\nu\sigma}+\theta_{\mu\sigma}\omega_{\nu\rho}+\theta_{\nu\rho}\omega_{\mu\sigma}+\theta_{\nu\sigma}\omega_{\mu\rho}) ~, \\ [0.5em] \nonumber
P^{(0,w)}_{\mu\nu,\rho\sigma} &=& \omega_{\mu\nu}\omega_{\rho\sigma} ~, \qquad P^{(0,sw)}_{\mu\nu,\rho\sigma} = \dfrac{1}{\sqrt{2}}\theta_{\mu\nu}\omega_{\rho\sigma} ~, \qquad P^{(0,ws)}_{\mu\nu,\rho\sigma} = \dfrac{1}{\sqrt{2}}\omega_{\mu\nu}\theta_{\rho\sigma} ~.
\end{eqnarray}
They are instrumental in writing down the expansion of the different terms in the action at quadratic level. For instance, if we expand the Einstein-Hilbert term we get
\begin{equation}
\mathcal{L}^{(2)}_{EH} = \sigma \sqrt{-g} R = \dfrac{\sigma}{2} h^{\mu\nu} \left[ P^{(2)}_{\mu\nu,\rho\sigma} - P^{(0,s)}_{\mu\nu,\rho\sigma} \right] \Box h^{\rho\sigma} ~.
\end{equation}
In order to compute the propagator we need to add a term in the Lagrangian fixing the de Donder gauge,
\begin{equation}
\mathcal{L}_{\rm gf}=-\dfrac{1}{2\alpha}\partial_{\mu}(\sqrt{-g}g^{\mu\nu})\partial^{\lambda}(\sqrt{-g}g_{\lambda\nu}) ~,
\end{equation}
whose quadratic expansion can be written using the above projector operators as
\begin{equation}
\mathcal{L}^{(2)}_{\rm gf} = \dfrac{1}{2\alpha}h^{\mu\nu} \left[ \dfrac{1}{2}P^{(1)} + \dfrac{1}{2} P^{(0,s)} + \dfrac{1}{4} P^{(0,w)} - \dfrac{1}{2\sqrt{2}}(P^{(0,sw)} + P^{(0,ws)}) \right]_{\mu\nu,\rho\sigma} \Box h^{\rho\sigma} ~.
\end{equation}
The quadratic expansion of the Chern-Simons term reads \cite{Pinheiro}
\begin{equation}
\mathcal{L}^{(2)}_{CS} = \dfrac{1}{\mu}\varepsilon^{\lambda\mu\nu}\Gamma^{\rho}_{\lambda\sigma} \left( \Gamma^{\sigma}_{\rho\nu,\mu} + \dfrac{2}{3} \Gamma^{\sigma}_{\mu\tau} \Gamma^{\tau}_{\nu\rho} \right) = \dfrac{1}{2\mu} h^{\mu\nu} \left[ S^{(1)}_{\mu\nu,\rho\sigma} + S^{(2)}_{\mu\nu,\rho\sigma} \right] \Box h^{\rho\sigma} ~,
\end{equation}
where we have introduced the spin operators
\begin{eqnarray}
& & S^{(1)}_{\mu\nu,\rho\sigma} = \dfrac{1}{4}\Box(\varepsilon_{\mu\rho\lambda}\partial_{\nu}\omega^{\lambda}_{\sigma}+\varepsilon_{\mu\sigma\lambda}\partial_{\nu}\omega^{\lambda}_{\rho}+\varepsilon_{\nu\rho\lambda}\partial_{\mu}\omega^{\lambda}_{\sigma}+\varepsilon_{\nu\sigma\lambda}\partial_{\mu}\omega^{\lambda}_{\rho}) ~, \\[0.5em] \nonumber
& & S^{(2)}_{\mu\nu,\rho\sigma} = -\dfrac{1}{4}\Box(\varepsilon_{\mu\rho\lambda}\eta_{\nu\sigma}+\varepsilon_{\nu\rho\lambda}\eta_{\mu\sigma}+\varepsilon_{\mu\sigma\lambda}\eta_{\nu\rho}+\varepsilon_{\nu\sigma\lambda}\eta_{\mu\rho}) \partial^{\lambda} ~.
\end{eqnarray}
The graviton propagator in TMG can then be written as (fixing $\alpha = 1$):
\begin{eqnarray}
\mathcal{D}^{\rm TMG}_{\mu\nu\alpha\beta} &=& \dfrac{i4(-p^{2})}{\frac{(-p^{2})^3}{\mu} - \frac{(\sigma\mu)^2}{\mu}(-p^{2})^2} \left[ -\sigma\mu P^{(2)}_{\mu\nu,\alpha\beta} - \frac{1}{4} (\epsilon_{\mu\alpha\lambda}\theta_{\beta\nu} + \epsilon_{\mu\beta\lambda}\theta_{\alpha\nu} + \epsilon_{\nu\alpha\lambda}\theta_{\beta\mu} \right. \nonumber \\ [0.5em]
& & \left. + \epsilon_{\nu\beta\lambda}\theta_{\alpha\mu}) (ip^{\lambda}) \right] + \dfrac{4}{\sigma(-p^2)} \left[ P^{(1)}_{\mu\nu,\alpha\beta} - P^{(0,s)}_{\mu\nu,\alpha\beta} - \sqrt{2} \left( P^{(0,sw)}_{\mu\nu,\alpha\beta} + P^{(0,sw)}_{\alpha\beta,\mu\nu} \right) \right] ~.
\label{propTMG}
\end{eqnarray}
In the case of NMG, the quadratic expansion of the Lagrangian reads
\begin{equation}
\mathcal{L}_{K}^{(2)} = \dfrac{1}{4m^2} h^{\mu\nu} P^{(2)}_{\mu\nu,\rho\sigma} \Box^2 h^{\rho\sigma} ~,
\end{equation}
which leads to a gauge fixed graviton propagator of the form
\begin{equation}
\mathcal{D}^{\rm NMG}_{\mu\nu\alpha\beta} = \dfrac{im^2}{(-p^2)(-p^2+\sigma m^2)}P^{(2)}_{\mu\nu,\alpha\beta} + \dfrac{2i}{\sigma(-p^2)}\left[ P^{(1)}_{\mu\nu,\alpha\beta} - P^{(0,s)}_{\mu\nu,\alpha\beta} - \sqrt{2} \left( P^{(0,sw)}_{\mu\nu,\alpha\beta} + P^{(0,sw)}_{\alpha\beta,\mu\nu} \right) \right] ~.
\label{propNMG}
\end{equation}
These are the propagators used along the paper to compute the Eikonal scattering amplitudes. 


\end{document}